\documentclass[12pt]{iopart}
\usepackage{epsfig,graphicx}
\usepackage{amstext}
\usepackage{amssymb}
\usepackage{amsthm}
\usepackage{amsfonts}
\usepackage{latexsym}
\def\beq{\begin{equation}}{\it}
\def\eeq{\end{equation}}
\def\beqa{\begin{eqnarray}}{\it}
\def\eeqa{\end{eqnarray}}

\begin{document}
\title{Fragmented condensation in Bose-Hubbard trimers with tunable tunnelling}

\author{A. Gallem\'{\i}$^{1,2}$, M. Guilleumas$^{1,2}$, 
J. Martorell$^1$, R. Mayol$^{1,2}$, A. Polls$^{1,4}$ and B. Juli\'a-D\'{\i}az$^{1,3,4}$}
\address{$^1$ Departament d'Estructura i Constituents de la Mat\`{e}ria,\\
Facultat de F\'{\i}sica, Universitat de Barcelona, E--08028 Barcelona, Spain}
\address{$^2$ Institut de Nanoci\`encia i Nanotecnologia de la Universitat 
de Barcelona (IN$\,^2$UB), E--08028 Barcelona, Spain}
\address{$^3$ ICFO-Institut de Ci\`encies Fot\`oniques, Parc Mediterrani 
de la Tecnologia, E--08860 Barcelona, Spain}
\address{$^4$ Institut de Ci\`encies del Cosmos, 
Universitat de Barcelona, IEEC-UB, Mart\'i i Franqu\`es 1, E--08028
Barcelona, Spain}
\begin{abstract}
We consider a Bose-Hubbard trimer, i.e. an ultracold Bose gas 
populating three quantum states. The latter can be either different 
sites of a triple-well potential or three internal states of the 
atoms. The bosons can tunnel between different states with variable 
tunnelling strength between two of them.  
This will allow us to study; i) different geometrical configurations, i.e. 
from a closed triangle to three aligned wells and ii) a triangular 
configuration with a $\pi$-phase, i.e. by setting one of the tunnellings 
negative. By solving the corresponding three-site Bose-Hubbard Hamiltonian 
we obtain the ground state of the system as a function of the trap 
topology. We characterise the different ground states by means of 
the coherence and entanglement properties. For small repulsive 
interactions, fragmented condensates are found for the $\pi$-phase 
case. These are found to be robust against small variations of the 
tunnelling in the small interaction regime. A low-energy effective 
many-body Hamiltonian restricted to the degenerate manifold provides 
a compelling description of the $\pi$-phase degeneration and explains 
the low-energy spectrum as excitations of discrete semifluxon states. 
\end{abstract}

\maketitle

\section{Introduction}
\label{sec:intro}

It is well known that bosons at sufficiently low temperatures tend to form 
Bose-Einstein condensates (BECs), which essentially consist on the macroscopic 
population of a single-particle state~\cite{Dalfovo1999}. In absence of 
interactions, the macroscopically occupied state is the lowest energy state 
of the single-particle Hamiltonian. When the interatomic interactions are 
taken into account, and for sufficiently large number of atoms, the main 
effect is a broadening of the single-particle state, which can be accounted 
for by a mean-field Gross-Pitaevskii description. In the Onsager-Penrose 
picture, in a BEC there is only one eigenvalue of the one-body density matrix 
which is of the order of the total number of particles. This is the largest 
eigenvalue and is termed the ``condensed fraction''. 

In contrast, an interesting scenario appears when, in absence of interactions, 
the lowest single-particle states are degenerate. In this case, the many-body 
ground state may get fragmented~\cite{Mueller2006}, as naively the atoms have 
no reason to condense in only one of the degenerate single-particle states. 
This implies that a finite number of eigenvalues of the single-particle density 
matrix are of order of the total number of atoms. 

In Ref.~\cite{Mueller2006} 
the authors describe three notable physical examples which produce fragmented 
condensates, e.g. the highly correlated regime in ultracold gases subjected
to synthetic gauge fields, the two-site Hubbard model and the ground state 
of a spinor condensate in absence of quadractic Zeeman terms.  
In the first two cases, interactions need to dominate over tunnelling terms 
in order to get fragmentation. For instance, for atoms in the 
double-well it is in the strongly 
repulsive regime that the condensate fragments in two parts, with half the 
atoms populating each well. By increasing the potential barrier 
between the wells, the system enters into the Fock regime (interaction energy dominating 
over the tunnelling) and fragments in two BECs of equal number of particles. 
This has been realized experimentally~\cite{Albiez2005,Levy2007,Gross2010,Riedel2010}. 
Notably, the quantum many-body correlations present in fragmented states can 
find applications in the field of quantum metrology to improve precision 
measurements~\cite{Gross2010}, and hold promise of being useful in near future 
technological applications~\cite{Nshii2013}.

It is desirable to pin down quantum many-body systems which feature fragmentation 
even at the single-particle level without explicit spatial separation. In 
this article we consider a minimal system which fulfills this, and which therefore 
has fragmented ground states both in absence of atom-atom interactions or for 
small ones. We consider $N$ identical bosons populating three single-particle 
states. The bosons are assumed to be able to tunnel between the different states. 
There are two options that are available with current techniques. The first 
one would be to trap the atoms in a triple-well potential, with fully connected 
sites as in~\cite{Nemoto2000,Lee2006}, or aligned~\cite{Viscondi2010}. In this case, the 
three quantum states are the three eigenstates of the single-particle Hamiltonian, 
which are thus spatially localised. A second option is to consider three different 
internal states of the boson as single-particle states, with the whole cloud 
being trapped on the same harmonic potential. Transitions among the three 
internal states can be induced by means of Rabi coupling. In this case, the 
three quantum states are not spatially localised. These two cases may be referred to as 
external or internal, three-mode systems. The properties of such triple-well 
potentials have been studied previously exploring the effect of dipolar 
interaction in the system~\cite{Lahaye2010,Mazzarella2013,Dell'Anna2013,Gallemi2013,xiong2013}, 
the many-body properties of the system~\cite{Viscondi2010,Viscondi2011} or 
the melting of vortex states~\cite{Lee2006}. 

In the previous studies, the coupling between the different modes is provided by 
the quantum tunnelling between the spatially localised modes. This has hindered 
the exploration of the regime we discuss in the present article. Namely, we will 
consider triangular setups in which one of the tunnelling terms can be taken 
negative. This means that a particle tunnelling from one site to the next one 
acquires a phase of $\pi$. Thus, we consider cases when a particle acquires
either 0 or $\pi$ phase when tunnelling around the triangle in absence of 
interactions. The key idea is to consider systems where tunnelling can be 
detuned~\cite{Tarruell2012} and particularly profit from the recent advances 
in producing phase dependent tunnelling terms. For external 
modal configurations, an external shaking of the system along one direction effectively 
results in a dressed tunnelling term whose sign can be switched from positive 
(standard) values to negative ($\pi$-phase tunnelling)~\cite{Eckardt2005}. 
More recently, a deep laser dip in the centre of a junction has been 
proposed in Ref.~\cite{Szirmai2014} to engineer $\pi$-phase tunnelling.
In the internal case a phase-dependent tunnelling can be obtained as in 
Ref.~\cite{Jaksch2003}. 

\begin{figure}[t]
\centering
\includegraphics[width=0.7\columnwidth]{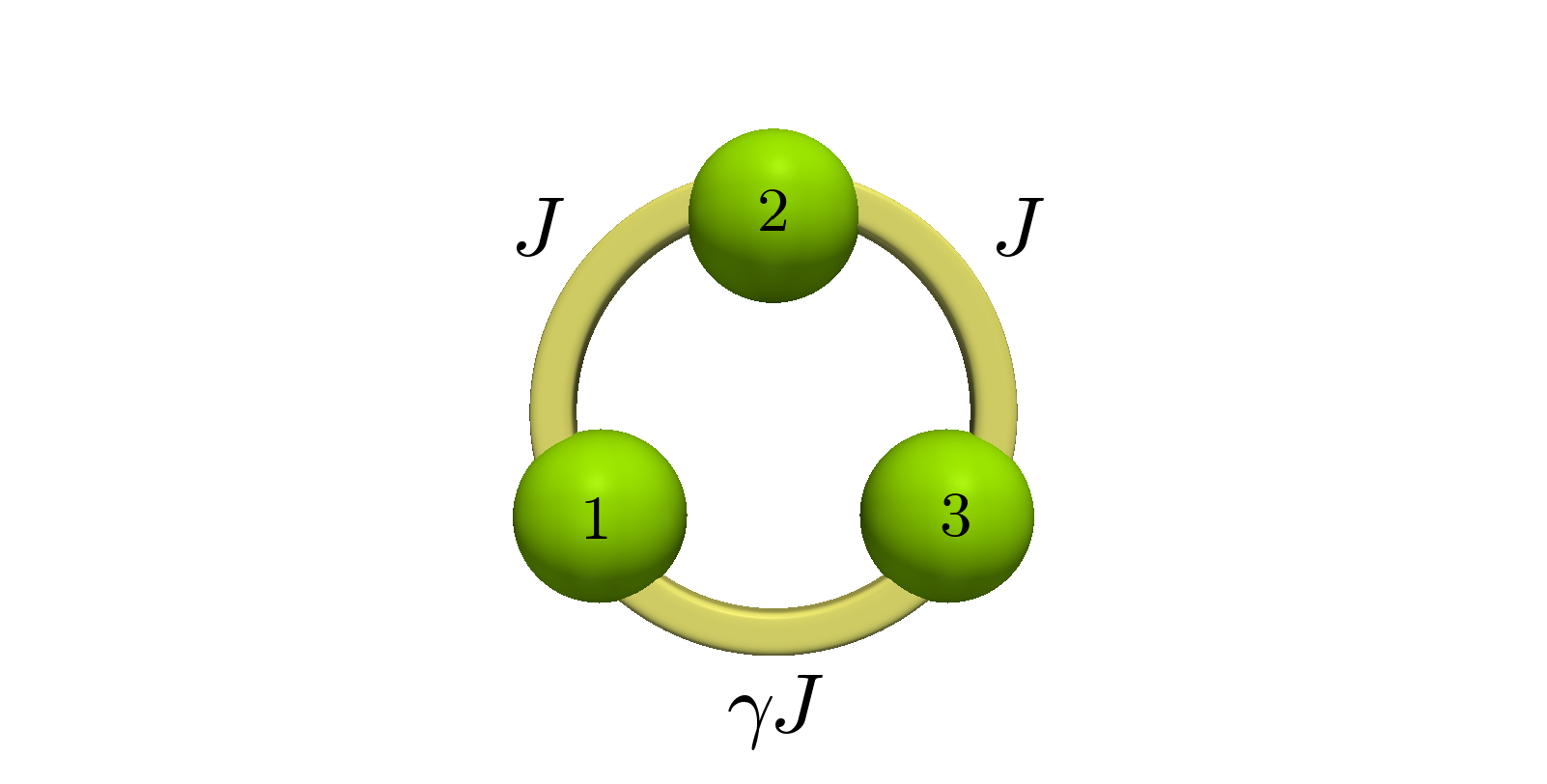}
\caption{Schematic depiction of the system under study. We consider a 
gas of bosons which can populate three different modes, depicted as 
green balls. Bosons are allowed to tunnel between the modes. The tunnelling 
between modes 1 and 3 is taken to be tunable through the parameter 
$\gamma$. For $\gamma=0$ the three sites are aligned and 
the geometry is essentially one dimensional. For $\gamma>0$ we have 
a triangular configuration, which for $\gamma=1$ is equilateral. In 
the limit of $\gamma>> 1$ the system is similar to a two-mode one. 
For $\gamma<0$ we have $\pi$-phase tunnelling between sites 1 and 3.}
\label{configurations}
\end{figure}

This paper is organised as follows. In Sec.~\ref{sec:bh}, we present the 
three-mode Bose-Hubbard Hamiltonian, and briefly recall the many-body 
magnitudes that are used to characterise the entanglement and correlation 
properties in the system. In Sec.~\ref{section-sp}, we first discuss the 
non-interacting case, whose properties are crucial to understand the 
onset of fragmentation at the many-body level. Sec.~\ref{section-frag} is 
devoted to analyse the role played by interactions. In Sec.~\ref{gam1} we 
provide a model developed to understand the appeareance of fragmentation 
in the symmetric $\pi$ phase. Finally, in Sec.~\ref{sec:con} we summarise 
our conclusions and provide possible implications for future experiments.

\section{Three-mode Bose-Hubbard Hamiltonian}
\label{sec:bh}

We assume $N$ ultracold bosons populating three quantum states. As discussed 
above, they can be different sites, e.g. an ultracold gas confined in a 
triple-well potential, or three internal states of the atoms. For the time 
being, we will restrict our system to the former case. We consider tunnelling 
terms between the three sites, and a tunable rate between two of them. 
This tunable rate will allow us to explore colinear configurations, closed 
ones, and also configurations with $\pi$ phase. Besides the tunnelling, 
the atoms interact via $s$-wave contact interactions. This interaction is assumed 
to be the same between all atoms, which is straightforward in the external 
case, as there is only one kind of atoms, but would require some tuning 
in the internal case depending on the species. The three-mode Bose-Hubbard 
(BH) Hamiltonian we consider is thus, 

\begin{eqnarray}
\hat{\mathcal{H}}&=&
-J\!\left[\hat{a}_1^\dagger \hat{a}_2 + \gamma \, \hat{a}_1^\dagger \hat{a}_3
+ \hat{a}_2^\dagger \hat{a}_3 + h.c. \right] +
\frac{U}{2} \sum_{i=1}^{3} \hat{n}_i (\hat{n}_i-1)  \,,
\label{hamiltonian}
\end{eqnarray}
where $\hat{a}_i$ ($\hat{a}_i^\dagger$) are the bosonic annihilation (creation) 
operators for site $i$ fulfilling canonical commutation relations, and 
$\hat{n}_i= \hat{a}_i^\dagger \hat{a}_i$ is the particle number operator on 
the $i$-th site. $J$ is the tunnelling coefficient between sites $1$-$2$, and 
$2$-$3$, and $U$ is the atom-atom on-site interaction that can be repulsive 
$U>0$ or attractive $U<0$. In our study, sites $1$ and $3$ are always equivalent 
with respect to site $2$, and the tunnelling between sites $1$-$3$ depends 
on the particular configuration through the parameter $\gamma$.

Figure~\ref{configurations} shows schematically the triple-well potentials 
we have addressed. When $\gamma=0$, no tunnelling exists between sites $1$ 
and $3$, which corresponds to an aligned triple-well configuration. When 
$0<\gamma < 1$, sites $1$ and $3$ are connected but the tunnelling 
between $1$-$2$ (and $2$-$3$ as well) is larger. In contrast, when $\gamma=1$, 
the three sites are fully equivalent with the same tunnelling rate among them, 
which can be geometrically interpreted as arranged in an equilateral 
triangular potential. We will go beyond this symmetric configuration, 
when $\gamma > 1$, by increasing the tunnelling rate between sites $1$ 
and $3$ with respect to $1$-$2$ (and $2$-$3$), up to $\gamma=2$ and we 
will consider negative values of $\gamma$ besides. The last range can be 
engineered by lattice shaking along the direction of sites $1$ and 
$3$~\cite{Eckardt2005}. 

The Hamiltonian of Eq.~(\ref{hamiltonian}) can also be reproduced with a 
three-component (spinorial, isotopic or atomic) BEC mixture trapped on a 
single harmonic oscillator with suitable Rabi coupling between the levels 
(which corresponds to the tunnelling terms).

Our study will concentrate mostly on the repulsive interaction case, 
which is more prone to be experimentally explored with current setups. 
We will also discuss briefly the attractive interaction case. 
In the latter even small asymmetries in the external trapping potentials 
will eventually have a large impact on the properties of the ground 
state~\cite{JuliaDiaz2010b}. Thus, we will introduce small symmetry breaking 
terms, compared to both the tunnelling and interaction, to consider 
situations closer to experimental ones and also to control numerically 
the degeneracies in the problem. We have added the biases between sites 
$1$-$3$ and $2$-$3$, $\epsilon_{13}$ and $\epsilon_{23}$, respectively. 
The considered Hamiltonian now reads:

\begin{equation}
\hat{\mathcal {H'}}= 
\hat{\mathcal{H}} 
+ \epsilon_{13}( \hat{n}_1 -\hat{n}_3 ) + \epsilon_{23}( \hat{n}_2 -\hat{n}_3 ) \,.
\label{bias}
\end{equation}

\subsection{Many-body basis}

The system is studied through direct diagonalisation of the many-body 
Hamiltonian for a fixed number of atoms, $N$. A suitable many-body basis 
is the Fock one, which labels the number of atoms in each mode,

\begin{equation}
|n_1,n_2,n_3\rangle = {1\over \sqrt{  n_1!\, n_2!\, n_3!}} 
\left( \hat{a}_1^\dagger\right)^{n_1}
\left( \hat{a}_2^\dagger\right)^{n_2}
\left( \hat{a}_3^\dagger\right)^{n_3} \ |{\rm vac}\rangle\,,
\end{equation}
where  $\vert {\rm vac} \rangle$ stands for the vacuum, and $N=n_1+n_2+n_3$. 
The elements of the Fock basis can be expressed as a product state 
$ |n_1,n_2,n_3\rangle = |n_1\rangle \otimes |n_2\rangle \otimes |n_3\rangle$.
A general many-body wavefunction is thus written as

\begin{equation}
\vert \Psi \rangle = \sum_{n_1,n_2}^N \, C_{n_1,n_2} \, \vert n_1,n_2,n_3\rangle\,,
\label{psifock}
\end{equation}
where $C_{n_1,n_2}$ is the corresponding amplitude of the Fock state 
$\vert n_1,n_2,n_3\rangle$, and $n_3=N-(n_1+n_2)$.

\subsection{Coherent states}

Coherent states are the closest analogs to classical 
solutions, in the same way wavepackets are the closest quantum analog 
to classical trajectories. A general coherent state can be constructed 
by assuming that all $N$ atoms populate the same single-particle state $\ell$,

\begin{equation}
\hat{b}_\ell^\dagger \equiv 
c_1 \, \hat{a}_1^{\dagger} + c_2 \, \hat{a}_2^{\dagger}+ c_3 \, \hat{a}_3^{\dagger} \,.  
\end{equation}
The coherent state reads, 

\begin{equation}
\vert \Psi_{\rm COH} \rangle \! 
=  \!\! \frac{1}{\sqrt{N!}}\left( \hat{b}_\ell^\dagger\right)^N |{\rm vac}\rangle\,.
\label{coherent-general}
\end{equation}
Since $c_i \in \mathbb{C}$, this many-body state has 6 parameters to be 
determined. Properly normalising the single-particle wavefunction and also 
realising that there is always an arbitrary global phase, the number 
of free parameters can be reduced to 4. It is worth stressing that a 
coherent state as the one defined above corresponds to a fully condensed 
atomic cloud. 

\subsection{Condensed fractions}

The fragmentation properties 
of the ultracold atomic gas~\cite{Mueller2006} can be investigated by 
means of the eigenvalues of the one-body density matrix. In our system 
with $N$ bosons and three different modes, the one-body density matrix 
of a many-body state $|\Psi\rangle$ is a $3 \times 3$ matrix whose elements 
are,

\begin{equation}
\hat{\rho}_{ij}^{(1)}={1\over N} \langle \Psi \vert \hat{a}_i^\dagger \hat{a}_j \vert \Psi \rangle\,,
\end{equation}
with $i,j=1,2,3$. Since $| \Psi \rangle$ is normalised, it follows that 
${\rm Tr} \, \hat{\rho}^{(1)}=1$. 

It is interesting to calculate its eigenvectors (natural 
orbitals), $|\psi_i\rangle$, and eigenvalues, $p_i$, with 
$p_1\geq p_2 \geq p_3 \geq 0$, that satisfy $p_1+p_2+p_3=1$. Each eigenvalue 
of the one-body density matrix gives the relative occupation number 
of the corresponding natural orbital: $p_i=N_i/N$. In a singly 
condensed system, there is only one large eigenvalue that corresponds 
to the condensed fraction of the single-particle state, $|\psi_1\rangle$: $p_1 \sim 1$ and 
$N_1 \sim {\cal O}(N)$, with all the other eigenvalues $p_j$ ($j\neq1$) being 
small $\sim {\cal O}(1/N)$. Instead a fragmented system has more than 
one large eigenvalue, $N_i \sim {\cal O}(N)$, with $i=1,\dots, s$,
and the rest of eigenvalues $p_j$ ($j>s$) are small $\sim {\cal O}(1/N)$. 
In this situation the system is not fully condensed, but fragmented, 
quantum correlations become important and the mean-field approximation 
fails to describe the system.

\subsection{Entanglement properties: entanglement entropy and Schmidt gap}
\label{sec:ent}

Correlations between different subsystems of a many-body quantum system 
can be quantified performing different bipartite splittings. That is, 
considering the system as made of two subsystems, tracing out one of 
the parts, and studying the von Neumann entropy and the entanglement 
spectrum~\cite{Li2008} of the resulting subsystem. 

In our case, we consider different spatial partitions of the three-well 
configuration, e.g. (1,23), (2,13), as in Ref.~\cite{Gallemi2013}. From the density matrix of the full 
system, $\hat\rho$, correlations between mode $i$ and the rest can be 
determined by first taking the partial trace of $\hat\rho $ over the 
Fock-state basis of the other modes. This yields the reduced density 
matrix on subsystem $i$, $\hat\rho_i$, that describes the state of this 
subsystem. For instance, in our system, by tracing out sites $2$ and $3$, 
a bipartite splitting of the three-mode system is obtained $(1,23)$, 
and the reduced density matrix on site $1$, 
$\hat{\rho}_1 = {\rm Tr}_{23} \hat \rho$, is found to be diagonal in the 
single mode space of $N$ particles (see Eq.~(\ref{rho_i}) in 
Appendix~\ref{app:diagonality}), 

\begin{equation}
\hat{\rho}_1= \sum_{k=0}^N \lambda^1_k |k\rangle\langle k| \ , \quad \lambda^1_k\geq 0
\label{eq:r1}
\end{equation}
where $|k\rangle$ are states of $k$ particles in mode 1. Note, that the 
reduced density matrix for state $1$ is in general a mixture without a 
well-defined number of particles. The set of eigenvalues $\{ \lambda_k^1 \} $ 
is called entanglement, or Schmidt spectrum~\cite{Li2008,DeChiara2012,footnote2}, 
and the eigenvalues are the Schmidt coefficients. The Schmidt coefficient 
$\lambda^1_k$ is in this case directly the probability of finding $k$ 
particles in site 1 without measuring the number of atoms in sites 2 and 3. 
The Schmidt spectrum fulfills $\sum_i \lambda_{i}^1 =1 $, and contains 
information about the correlations and the entanglement properties 
of the state in subsystem $1$ with respect to the rest of the system. 
It is worth recalling that a many-body state is entangled when it cannot 
be written as a product state. In the case of spatially separated 
modes, the entanglement we are discussing is spatial.

A measure of the entanglement between the two subsystems is 
already provided by the single-site von Neumann entropy, which 
can be computed as $S_1= - {\rm Tr}(\hat{\rho}_1 \log \hat{\rho}_1)$. 
Noting that in our case the density matrix, ${\hat{\rho}_1}$, is already diagonal, 
see Eq.~(\ref{eq:r1}), the entropy can be evaluated from 
the Schmidt coefficients as $S_1=-\sum_{i} \lambda_i^1 \log \lambda_i^1$. 

In the three-mode system, a signature of the entanglement on the Schmidt 
spectrum is the following one: if site $1$ is not entangled with sites 
$2$ and $3$, it should be pure after tracing those sites out, and then 
the entanglement spectrum would be $\lambda_1^1=1$ and $\lambda_i^1=0$ 
for $i=2,3, \dots$. This actually implies a zero of the corresponding 
von Neumann entropy. A remarkable magnitude defined from the set of 
$\lambda$'s is the so-called Schmidt gap, defined as the difference 
between the two largest and more relevant Schmidt coefficients in 
the entanglement spectrum of subsystem $i$, 
$\Delta{\lambda}^{i}=\lambda_{1}^{i}-\lambda_{2}^{i}$~\cite{DeChiara2012}. 
In the case of no entanglement between the subsystems, the Schmidt 
gap takes its maximum value of 1. A vanishing of the Schmidt gap 
marks large entanglement between the subsystems. As has been 
recently pointed out in Ref.~\cite{Gallemi2013} for dipolar bosons 
in triple-well potentials, the Schmidt gap is a good tool to distinguish 
between phase transitions and crossovers.

Note that no relation exists between fragmentation and entanglement, 
in the sense that a system can be entangled and fragmented independently. 
For instance, a system where all the bosons occupy the same spatial 
mode is nor fragmented neither spatially entangled. A bosonic Josephson 
junction in the strong repulsive interaction (Fock) regime is a clear 
example of a non entangled but fragmented state: $| N/2, N/2 \rangle$. 
In contrast, in the non-interacting regime the bosonic Josephson junction 
is in a fully condensed state, 

\beq
|\Psi\rangle = {1\over\sqrt{N!}}  \left({\hat{a}_1^\dagger+\hat{a}_2^\dagger
\over \sqrt{2}}\right)^N |{\rm vac}\rangle\,,
\eeq
which has large entanglement between the two sites as seen by the 
Schmidt coefficients which are $\lambda_k = 2^{-N} {N\choose k}$. 
Finally, cat states, $|\Psi\rangle =(|N, 0 \rangle + |0, N \rangle)/\sqrt{2}$, 
are a well known example of fragmented and entangled many-body systems.

\begin{figure}[t]
\centering
\includegraphics[width=0.7\columnwidth]{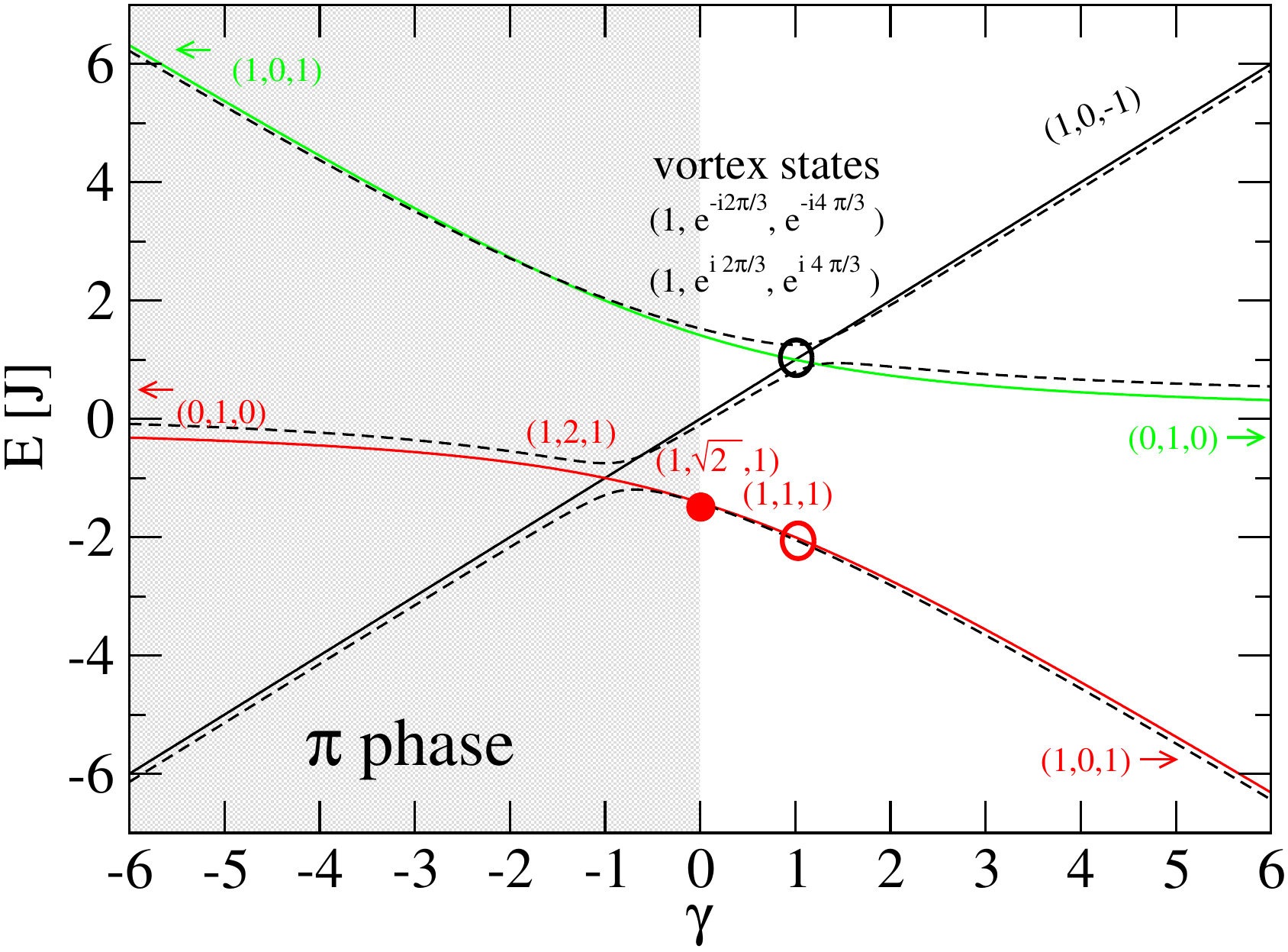}
\caption{Single-particle spectrum of the triple-well system as 
a function of $\gamma$. In black, red and green solid lines we 
depict the eigenvalues $E_1$, $E_2$ and $E_3$, from Eqs.~(\ref{sp:e}). 
The corresponding eigenstates, Eq.~(\ref{sp:eig}), are written 
explicitly for several relevant values of $\gamma$. 
The dashed lines correspond to the spectrum computed with a bias term 
$\epsilon_{13}=\epsilon_{23}=J/4$. The short notation for states $(a,b,c)$ in the 
figure corresponds to the states $a|1,0,0\rangle+b|0,1,0\rangle+c|0,0,1\rangle$ 
used in the main text.}
\label{singlep}
\end{figure}

\section{Non-interacting case}
\label{section-sp}

In this section we solve the non-interacting case for any value of the 
tunable tunnelling link $\gamma J$. In our calculation we will keep the 
tunnelling parameter fixed $J/h=1$ Hz, since it essentially sets the overall 
energy scale. 

The triple-well configuration we have chosen is intended to remark
first, the role played by the topology of the configuration, going from a 
disconnected chain ($\gamma=0$) to an essentially double-well system at 
$\gamma \to \infty$, through a connected equilateral triangle at $\gamma=1$. 
It is worth emphasising that we do so by varying the tunnelling strength 
between sites 1 and 3 and not by, for instance, adding sizeable bias terms 
to the Hamiltonian, which would indeed also break the symmetry between 
the three sites. The second important point, is that we consider 
$\pi$-phase tunnelling between the two wing sites ($\gamma<0$), which 
will indeed have dramatic consequences on the many-body properties 
of the system.

\subsection{Unbiased non-interacting case}

The non-interacting case can be solved readily. The eigenvalues of the Hamiltonian 
(\ref{hamiltonian}) with $U=0$ are,

\begin{eqnarray}
E_1&=&\gamma J \nonumber \\
E_2&=& {J\over 2}\left( -\gamma -\sqrt{8+\gamma^2}\right)\nonumber \\
E_3&=& {J\over 2}\left( -\gamma +\sqrt{8+\gamma^2}\right) \,,
\label{sp:e}
\end{eqnarray}
and the corresponding unnormalized eigenvectors are: 

\begin{eqnarray}
|\psi_1\rangle &=& {1\over\sqrt{2}}\left(|1,0,0\rangle -|0,0,1\rangle \right)\nonumber\\
|\psi_2\rangle &=& |1,0,0\rangle +
{4-\gamma^2+\gamma\sqrt{8+\gamma^2}\over \sqrt{8+\gamma^2}+3\gamma}\ |0,1,0\rangle 
+  |0,0,1\rangle \nonumber\\
|\psi_3\rangle &=& |1,0,0\rangle + 
{\gamma^2+\gamma\sqrt{8+\gamma^2}-4\over \sqrt{8+\gamma^2} - 3\gamma} 
\ |0,1,0\rangle + |0,0,1\rangle  \,.
\label{sp:eig}
\end{eqnarray}

The single-particle spectrum, see Fig.~\ref{singlep}, has some interesting 
features. First, there is one eigenvector that is independent of 
$\gamma$, $|\psi_1\rangle$. This state does not involve site 2, and 
its energy is proportional to the coupling between sites 1 and 3. 
For $\gamma=0$, which corresponds to an aligned configuration, the 
ground state is
$|\psi_2\rangle_{\gamma=0}= {1/2} \left(|1,0,0\rangle 
+ \sqrt{2} \,|0,1,0\rangle+|0,0,1\rangle \right)$, 
which has an excess of particles in site $2$. 

As $\gamma$ is increased, the equilateral triangular configuration is reached 
for $\gamma=1$. This case has been studied thoroughly in Refs.~\cite{Lee2006,Casetti02}.
The ground state is 
$|\psi_{2}\rangle_{\gamma=1}= (1/\sqrt{3})(|1,0,0\rangle + |0,1,0\rangle+|0,0,1\rangle)$ 
and there are two excited degenerate single-particle states that correspond 
to vortex states, with clockwise and counter-clockwise rotation:

\begin{eqnarray}
|\psi_{v1}\rangle &=& {1\over\sqrt{3}} 
\left(|1,0,0\rangle + e^{i{ 2\pi\over 3}} |0,1,0\rangle 
+ e^{i{ 4\pi\over 3}} |0,0,1 \rangle\right) \nonumber\\ 
|\psi_{v2}\rangle &=& {1\over\sqrt{3}}
\left(
|1,0,0\rangle 
+ e^{-i{ 2\pi \over3}} |0,1,0\rangle 
+ e^{-i{ 4\pi\over3}} |0,0,1\rangle\right) \,.
\label{vortexantivortex}
\end{eqnarray}
Notice that these two states are linear combination of 

\begin{eqnarray}
|\psi_1\rangle &=& {1\over\sqrt{2}}\left( 
e^{-i{\pi\over6}}|\psi_{v1}\rangle +
e^{i{\pi\over6}}|\psi_{v2}\rangle \right)
 \nonumber \\
|\psi_3\rangle_{\gamma=1} &=& 
{1\over\sqrt{6}}\left(|1,0,0\rangle -2 |0,1,0\rangle + |0,0,1\rangle\right) 
\nonumber \\
&=& {1\over\sqrt{2}}\left( 
e^{i{\pi\over3}}|\psi_{v1}\rangle +
e^{-i{\pi\over3}}|\psi_{v2}\rangle \right)
  \,.
\end{eqnarray}
Further increasing $\gamma$, sites 1 and 3 get further connected and 
the physics decouples them from site $2$. The ground state for 
$\gamma \to \infty$ is 
$|\psi_{2}\rangle_{\gamma \to \infty}= {1/\sqrt{2}} \left(|1,0,0\rangle +|0,0,1\rangle \right)$, 
the first excited state is 
$|\psi_3\rangle_{\gamma \to \infty}= |0,1,0\rangle$ and the second excited state is 
$|\psi_1\rangle_{\gamma \to \infty}={1/\sqrt{2}} \left(|1,0,0\rangle -|0,0,1\rangle \right)$.

The situation for $\gamma<0$ is very different and actually richer at the ground-state 
level. For $\gamma=-1$ we have a crossing in the single-particle spectrum, which 
therefore should have important consequences at the many-body level. 
At $\gamma=-1$ the ground state is two-fold degenerate between states 
$|\psi_1\rangle$ and 
$|\psi_2\rangle_{\gamma=-1}= \left(|1,0,0\rangle + 2 |0,1,0\rangle + |0,0,1\rangle \right)/\sqrt{6} $. 
The first excited state is 
$|\psi_3\rangle_{\gamma=-1}=(|1,0,0\rangle - |0,1,0\rangle + |0,0,1\rangle)/\sqrt{3}$. 
When varying $\gamma$ from $\gamma<-1$ to $\gamma>-1$, the system undergoes a quantum phase 
transition in this non-interacting case. A possible order parameter would be the 
population of the second site, which goes from $\langle \hat{n}_2\rangle=0$ to 
$\langle \hat{n}_2\rangle>0$ as we cross the $\gamma=-1$ point. 

From Fig.~\ref{singlep}, one can see that the main effect of 
varying $\gamma$ 
is a large avoided crossing of the asymptotic states 
$(|1,0,0\rangle +|0,0,1\rangle)/\sqrt{2}$  and $|0,1,0\rangle$. When $\gamma$ is
 positive the ground state is 
$(|1,0,0\rangle +|0,0,1\rangle)/\sqrt{2}$, whereas when $\gamma$  is 
negative the lowest state is the antisymmetric state, 
$|\psi_1\rangle =\,(|1,0,0\rangle -|0,0,1\rangle)/\sqrt{2}$. 
The latter remains uncoupled for 
all values of $\gamma$, with eigenenergy $E= J \gamma$. This value is easy to obtain, 
since in the Fock basis:  
$[ a^{\dag}_1 a_3 + a^{\dag}_3 a_1 ] |\psi_1\rangle = -|\psi_1\rangle$. Therefore, 
$\langle \psi_1| -J \gamma (  a^{\dag}_1 a_3 + a^{\dag}_3 a_1 ) |\psi_1\rangle= J \gamma$.

\subsection{Explicit symmetry breaking, effect of the bias}

The explicit symmetry breaking induced by bias terms in the Hamiltonian (\ref{bias}) 
breaks the degeneracies present in the single-particle spectrum. As seen in 
Fig.~\ref{singlep}, the crossings at $\gamma=-1$ and $\gamma=1$, which occur 
in the non-interacting system for the ground state and excited states, respectively, 
become now avoided crossings. For the case we are interested in, when the bias 
is much smaller than the tunnelling, we can obtain the states dressed by the 
bias at the degeneracy points by means of first order perturbation theory. In the case of 
$\epsilon_{13}> 0$ and $\epsilon_{23}=0$, the ground-state manifold at 
$\gamma=-1$ splits, $\Delta E=\epsilon_{13}$, and the corresponding dressed 
states are, 

\begin{eqnarray}
&|\tilde{\psi}_1\rangle =
{1\over\sqrt{2}} \left(|\psi_2\rangle -|\psi_1\rangle\right)=\nonumber\\
&=\frac{1}{\sqrt{12}}\left[(1-\sqrt{3})|1,0,0\rangle
+2|0,1,0\rangle  + (1+\sqrt{3})|0,0,1\rangle \right] \,
\nonumber\\
&|\tilde{\psi}_2\rangle =
{1\over\sqrt{2}} \left(  |\psi_1\rangle+|\psi_2\rangle\right)\nonumber\\
&=\frac{1}{\sqrt{12}}\left[(1+\sqrt{3})|1,0,0\rangle
+2|0,1,0\rangle  + (1-\sqrt{3})|0,0,1\rangle \right] \,.
\end{eqnarray}
The first one is the new ground state of the system, which has a slightly 
larger population of particles in site 3. This result is reasonable, since 
the bias we have considered, $\epsilon_{13} >0$, promotes site $3$.

\section{Quantum many-body properties of the system}
\label{section-frag}

We consider now the effect of repulsive interactions between atoms. We 
calculate the ground state, by exact diagonalisation of the Hamiltonian 
(\ref{hamiltonian}), for different values of the tunnelling rate $\gamma$. 
In our numerics we will consider up to $N=48$ particles. The ground state 
of the system is characterised by means of, a) coherence properties and 
fragmentation, b) analysis of the energy spectrum, and c) entanglement 
spectrum and entanglement entropy.

\subsection{Analytic results in the $|\gamma|\gg 1$ limit}
\label{sec:sm}

The structure of the single-particle spectrum, see Fig.~\ref{singlep}, 
allows one to build a simple model for $|\gamma|\gg 1$. 
This simple model will serve as guidance to understand many of the 
properties of the many-body ground state which will be discussed in the 
forthcoming sections. 

\begin{figure}[t]
\centering
\includegraphics[width=0.8\columnwidth]{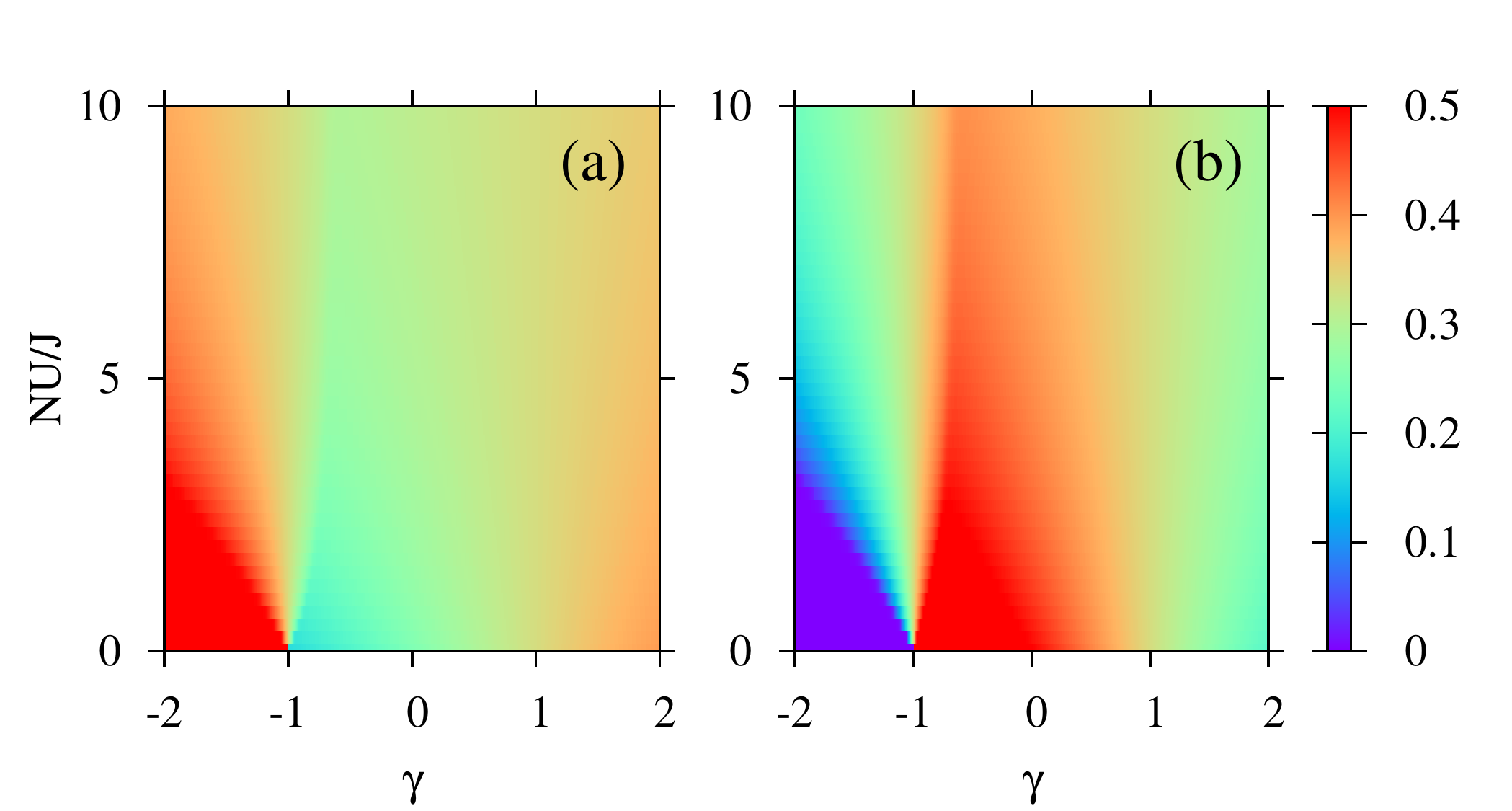}
\caption{Phase diagram in the ($\gamma,\,N U/J$) plane obtained by exact 
diagonalisation for $N=48$. The colour gives the average occupation in the 
ground state, $\langle \Psi_{\rm GS} |\hat{n}_i |\Psi_{\rm GS} \rangle /N$, 
of site $1$ (a) and $2$ (b). }
\label{diafases}
\end{figure}

We can distinguish two distinct regimes, a) $\gamma\gg1$ and 
b) $-\gamma\gg1$. In both cases, for sufficiently large values of 
$|\gamma|$ the single-particle spectrum approaches $E_0=-|\gamma| J$, 
$E_1=0$, and $E_2=|\gamma| J$. In the (a) case, the non-interacting 
single-particle states are 
$|\psi_0\rangle={1/\sqrt{2}}(|1,0,0\rangle + |0,0,1\rangle)$, 
$|\psi_1\rangle=|0,1,0\rangle$ and 
$|\psi_2\rangle={1/\sqrt{2}}(|1,0,0\rangle - |0,0,1\rangle)$, respectively. 
In (b) the single-particle states exchange their role, with $|\psi_2\rangle$ 
being the ground state and $|\psi_0\rangle$ the more energetic. For interaction 
strengths such that $NU/J \ll |\gamma|$, the effect of the interaction can be 
considered perturbatively. In this limit we can ignore particle-hole 
excitations to the highest single-particle state, and write configurations 
with $k$ excited atoms as, 

\begin{equation}
|\Psi\rangle= {\cal N} (\hat{a}^\dagger_{\psi_0})^{N-k} (\hat{a}^\dagger_{\psi_1})^k |{\rm vac}\rangle
\end{equation}
where ${\cal N}$ is a normalisation constant and $\hat{a}^\dagger_{\psi_i}$ creates 
a particle in the single-particle state $|\psi_i\rangle$. The energy for 
this state, up to constant terms, reads, 

\begin{equation}
E(N,k)= U {N-k\over2}\left({N-k+1\over2}-1\right) + {U \over2 } k(k-1) + |\gamma| J k\,.
\end{equation}
In the non-interacting case, the energy is minimal for $k=0$, as expected. 
For non-zero interactions, this expression has a minimum for 

\begin{equation}
k={\rm Int}\left[{N\over 3}\left( 1- {4|\gamma| J-U\over NU}\right)\right]\,,
\end{equation}
where the ${\rm Int}[\dots ]$ stands for the integer part of the expression. 
Incidentally, for large interactions, $NU/J\gg1$ (with $|\gamma|J\gg U$), 
the minimal energy is obtained 
for $k=N/3$, as expected for a Mott insulator of filling $N/3$. As 
we decrease interactions, the system goes step by step to $k=0$. 
In particular, many-body states with $k$ and $k+1$ excited atoms degenerate if, 
\begin{equation}
{NU\over J} = { 2 N |\gamma| \over N -3k -1} \,.
\label{eq:deg}
\end{equation}

\begin{figure}[t]
\centering
\includegraphics[width=.7\columnwidth]{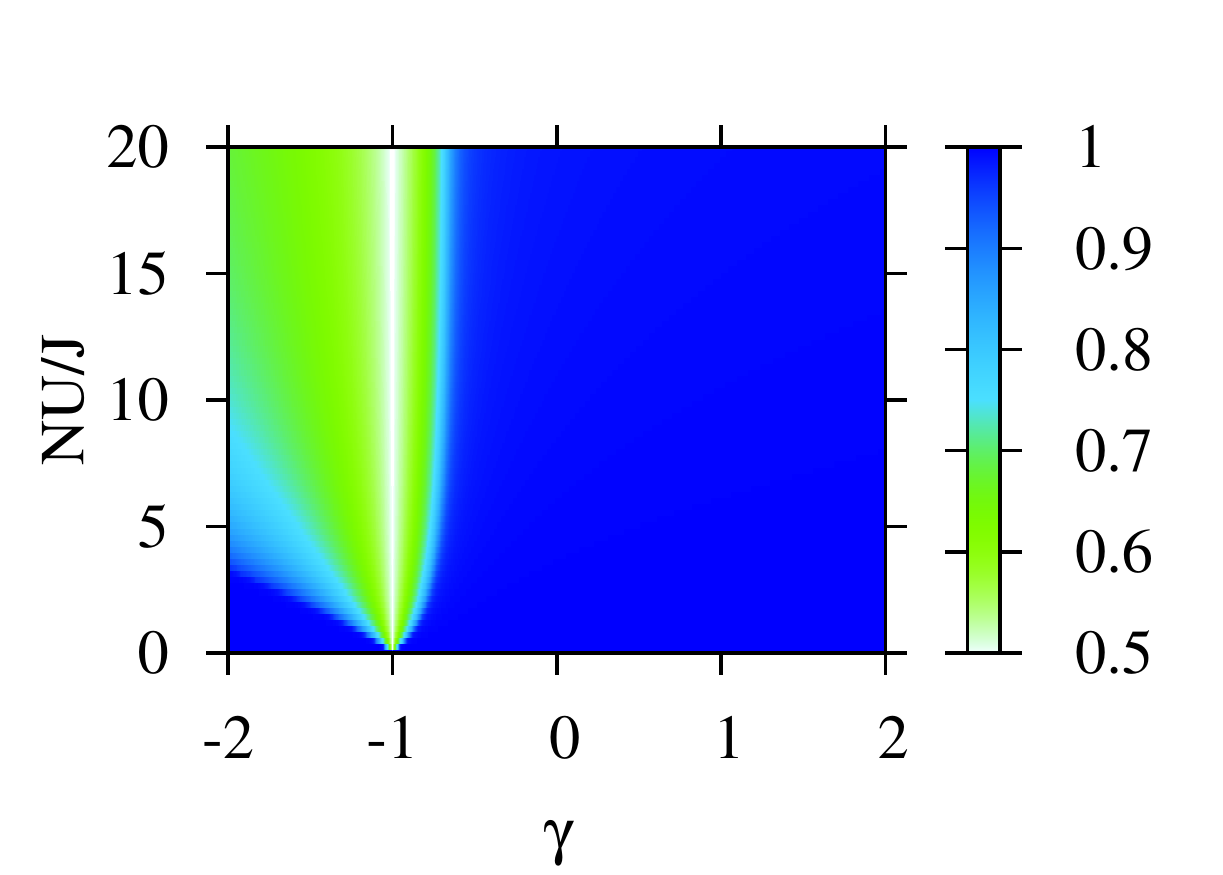}
\caption{Condensed fraction $p_1$ (largest eigenvalue of the one-body 
density matrix) as a function of $\gamma$ and $NU/J$. Notice that in 
this parameter
regime $p_1+p_2 \simeq 1$ and $p_3\simeq 0$. $N=48$. }
\label{nn}
\end{figure}
In this limit, it is quite reasonable to expect that as we increase the 
atom-atom interactions keeping $\gamma$ fixed, the system tends to 
minimise the number of pairs per site, which is achieved by equipopulating 
the three sites. Eq.~(\ref{eq:deg}) predicts ground state degeneracies, 
i.e. energy crossings, for certain values of $|\gamma|$ and $NU/J$. 

This model works reasonably 
both for $\gamma \lesssim -1$ and $\gamma \gtrsim 1$, where already the 
structure of the single-particle spectrum starts to resemble the asymptotic 
one. There is one important difference between $\gamma \lesssim -1$ and 
$\gamma \gtrsim 1$. In the former case, the lowest energy single-particle 
state is independent of $\gamma$ (see Fig.~\ref{singlep}) and has no 
population of site 2. This makes the model outlined above fairly accurate 
to describe the different transitions in the many-body ground state. In 
the latter however, the lowest single-particle state only asymptotically 
approaches the state $(1/\sqrt{2})(|1,0,0\rangle + |0,0,1\rangle)$. In 
this case, the model only provides a qualitative picture at smaller values of 
$|\gamma|$. Note also, that for $|\gamma|\gg1$, the value of $k$ which 
minimises the energy, corresponds in this limit to the average population 
of site $2$ in the ground state, while the average population of sites 
1 and 3 is $(N-k)/2$.

\begin{figure}[t]
\centering
\includegraphics[width=0.85\columnwidth]{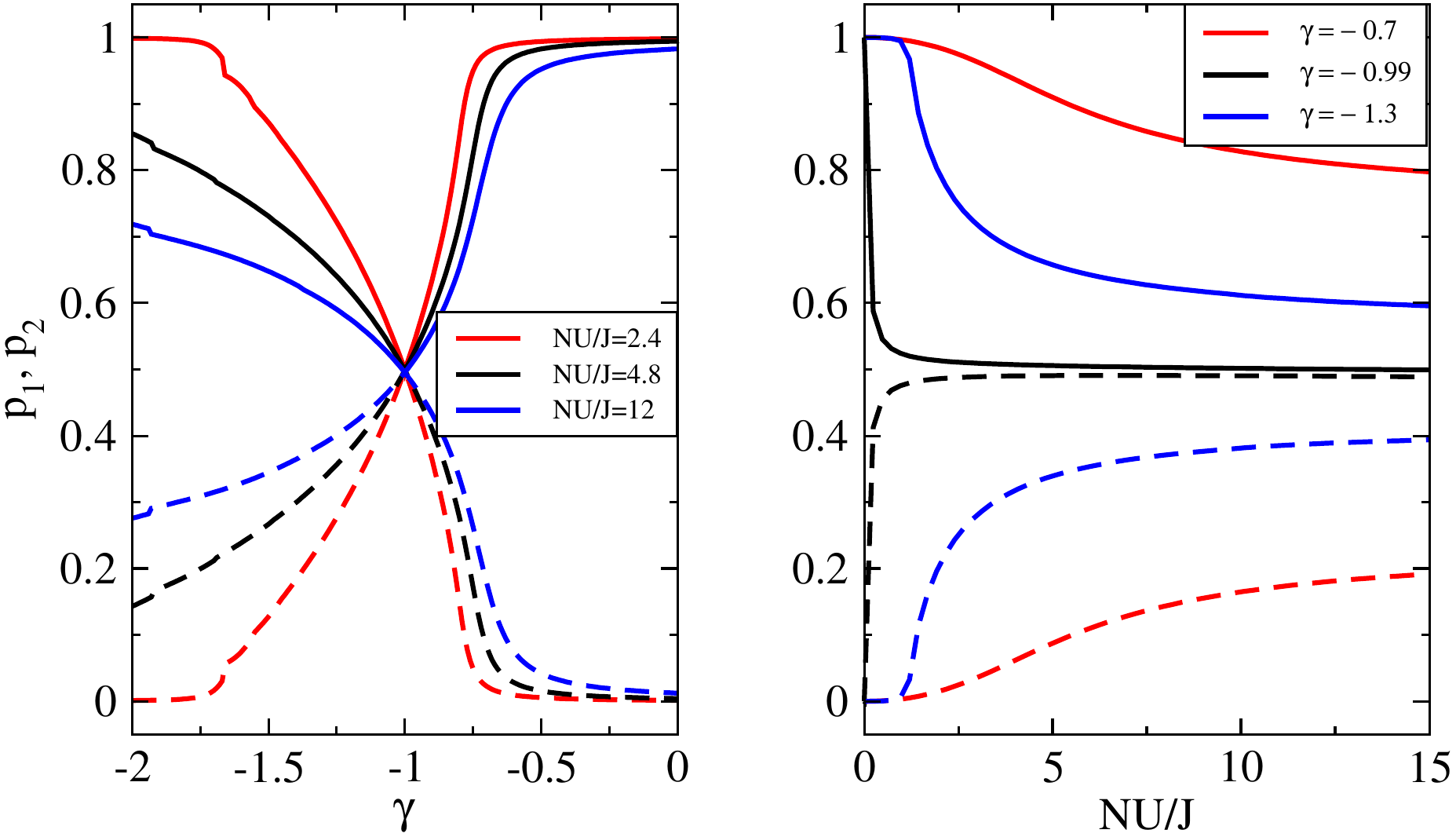}
\caption{The two largest eigenvalues of the one-body density matrix, 
$p_1$ (solid) $>$ $p_2$ (dashed), are depicted as a function of 
$\gamma$ (left) and $NU/J$ (right). }
\label{frag}
\end{figure}

\subsection{Coherence and one-body density matrix}

To have an overall picture of the different regimes that we will encounter 
for different values of $NU/J$ and $\gamma$ we first study the average 
populations of the three sites, $\langle \Psi_{\rm GS } | \hat{n}_i |\Psi_{\rm GS}\rangle/N$. 
Notice that in our system, without bias terms, sites $1$ and $3$ are equivalent.
In Fig.~\ref{diafases} we present results for $-2\leq \gamma \leq 2$ and 
relatively small values of the interaction $NU/J\leq 10$, which do not 
reach the Fock regime. For very small values of the dimensionless parameter 
$NU/J$, one does not expect substantial changes from the single-particle 
case. Indeed in the ground states for $\gamma > -1$ the three modes are 
substantially populated, whereas for $ \gamma < -1$ the second mode 
is clearly less occupied than the other two as reflected in the dark  
region on the left corner of Fig.~\ref{diafases} (b).

In the non-interacting limit, $U=0$, the problem becomes a single-particle one 
and in the case of a symmetric configuration, $\gamma=1$, the many-body ground 
state can be written as~\cite{Lee2006,Casetti02},
\begin{equation}
|\Psi_{\rm GS}^{U=0} \rangle = {1\over \sqrt{N!}}
\left({1\over \sqrt{3}}[
\hat{a}^\dagger_1+
\hat{a}^\dagger_2+
\hat{a}^\dagger_3]\right)^N \, |{\rm vac}\rangle \,,
\label{coh3}
\end{equation}
in which the average population on each site is, for symmetry reasons, $N/3$. 

For large repulsive atom-atom interactions, regardless of the value of $\gamma$, 
the system will fragment in an effort to diminish the number of pairs inside 
each site, in analogy to, for instance, the double-well~\cite{Mueller2006}. 
In the large interaction limit, that is in the Fock regime $U\gg J$, the ground 
state can be well approximated by 
$|\Psi_{\rm GS}^{U\gg J}\rangle \simeq |N/3,N/3,N/3\rangle\,$, which is the 
equivalent to a Mott insulator of filling $N/3$. This ground state is 
non-degenerate if the number of bosons is commensurate with $3$. If the number 
of bosons is not proportional to $3$ the ground state becomes three-fold degenerate 
in the strong Fock regime: $|N/3\pm1,N/3,N/3\rangle$, $|N/3,N/3\pm1,N/3\rangle$, 
and $|N/3,N/3,N/3\pm1\rangle$, where the plus (minus) sign refers to a single 
particle (hole) delocalization.

In the Josephson regime, defined as $NU/J\simeq 1$, the ground state 
of the system is mostly condensed for $\gamma\lesssim -0.5$. This is reflected 
in the eigenvalues of the one-body density matrix, see Fig.~\ref{nn} and 
left panel in Fig.~\ref{frag}. As already pointed out in the 
single-particle spectrum, a very interesting feature is readily 
found in the vicinity of $\gamma=-1$. In this case, the ground state 
of the system is fragmented in two pieces even in the non-interacting 
case. As the interaction increases, but still in the Josephson regime, 
the system is seen to remain bifragmented. From Fig.~\ref{frag}, one 
can see that the region in the $\gamma$-space, around $\gamma=-1$, 
which corresponds to a fragmented condensate, slightly increases with 
the interaction.

\begin{figure}[t]
\centering
\includegraphics[width=0.9\columnwidth]{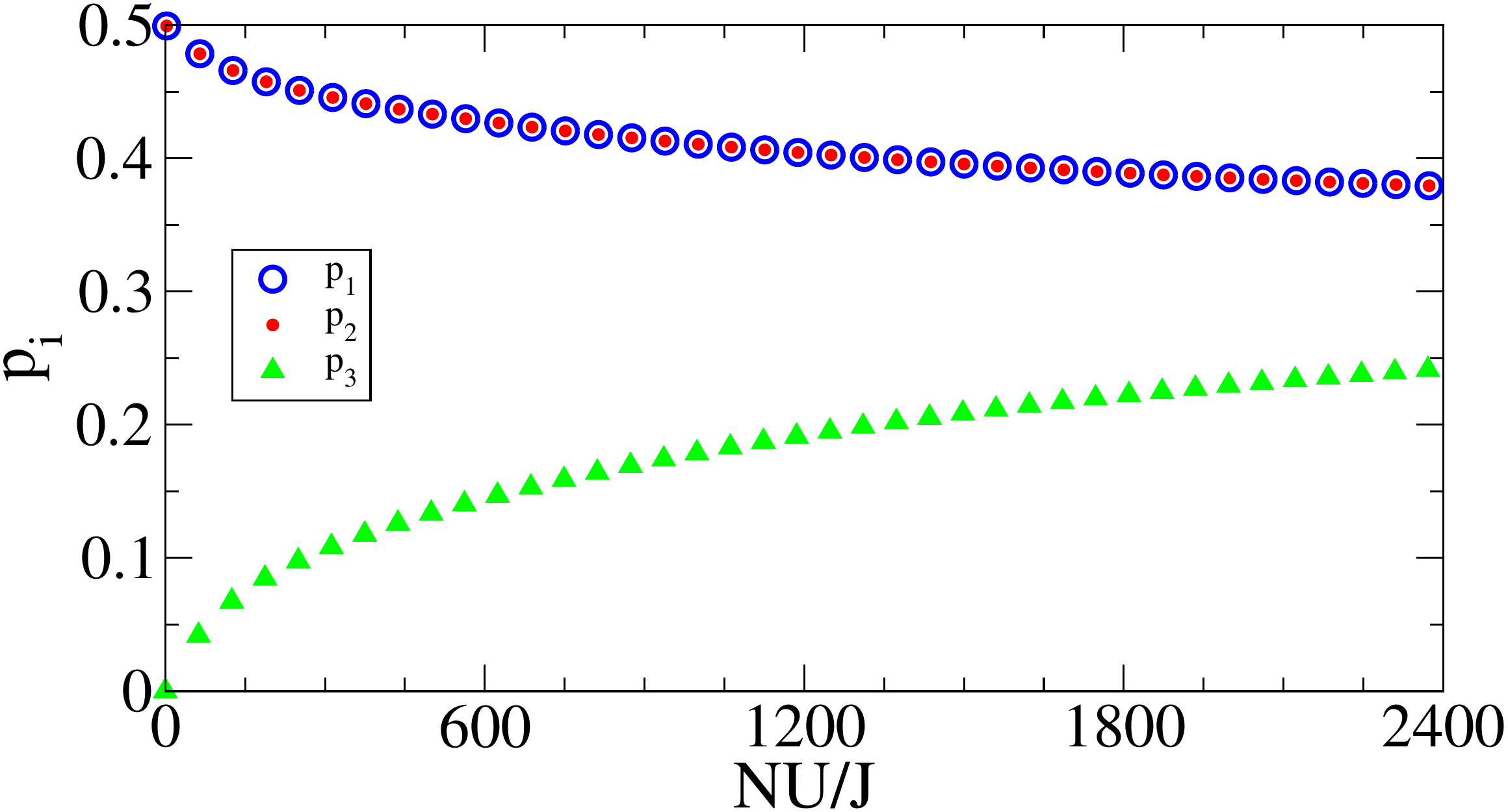}
\caption{Depiction of the three ordered eigenvalues of the one-body 
density matrix as a function of $NU/J$ for $\gamma=-1$. The figure 
shows the transition from bifragmentation in the $NU/J\simeq 1$ regime, 
to trifragmentation in the $NU/J\to \infty$ (Fock) regime.}
\label{cfrag}
\end{figure}

As discussed above, in the limit of the strong Fock regime, the 
ground state should be essentially fragmented in three pieces 
(in our case $N=48$, which is multiple of 3), corresponding to 
the ground state $|N/3,N/3,N/3\rangle$. This is clearly seen in 
Fig.~\ref{cfrag}, which depicts the behavior of the three eigenvalues 
of the one-body density matrix in the $\gamma=-1$ configuration 
for different values of the interaction. In the non-interacting case, 
the system is fragmented in two condensates ($p_1=p_2=0.5$ and $p_3=0$), 
whereas as $NU/J$ increases, $p_1=p_2 < 0.5$ decrease and $p_3>0$ 
increases, fulfilling $p_1+p_2+p_3=1$. Moreover, one can see that
$p_i \to 1/3$ asymptotically. Note however that the origin of 
bifragmentation is directly related to the degeneracy 
at the single-particle level and remains in the presence of 
tunnelling. Trifragmentation requires strong interactions 
such that essentially tunnelling plays no role and the system can 
be regarded as three independent condensates. 

In the right panel of Fig.~\ref{frag} the behavior of the two 
largest eigenvalues of the one-body density matrix is shown, as a 
function of $NU/J$, for a fixed $\gamma$ configuration. At 
$\gamma=-0.99$ there is a sharp transition from a singly condensed 
(at $U=0$) to a bifragmented system already for very small values 
$NU/J >0$. However, as $\gamma$ departs further from $-1$ ($\gamma=-0.7$ 
and $-1.3$) the transition between both regimes becomes smoother, 
and the system remains fully condensed for a larger 
range of interactions.

\subsection{Energy spectrum}

\begin{figure}[t]
\centering
\includegraphics[width=0.9\columnwidth]{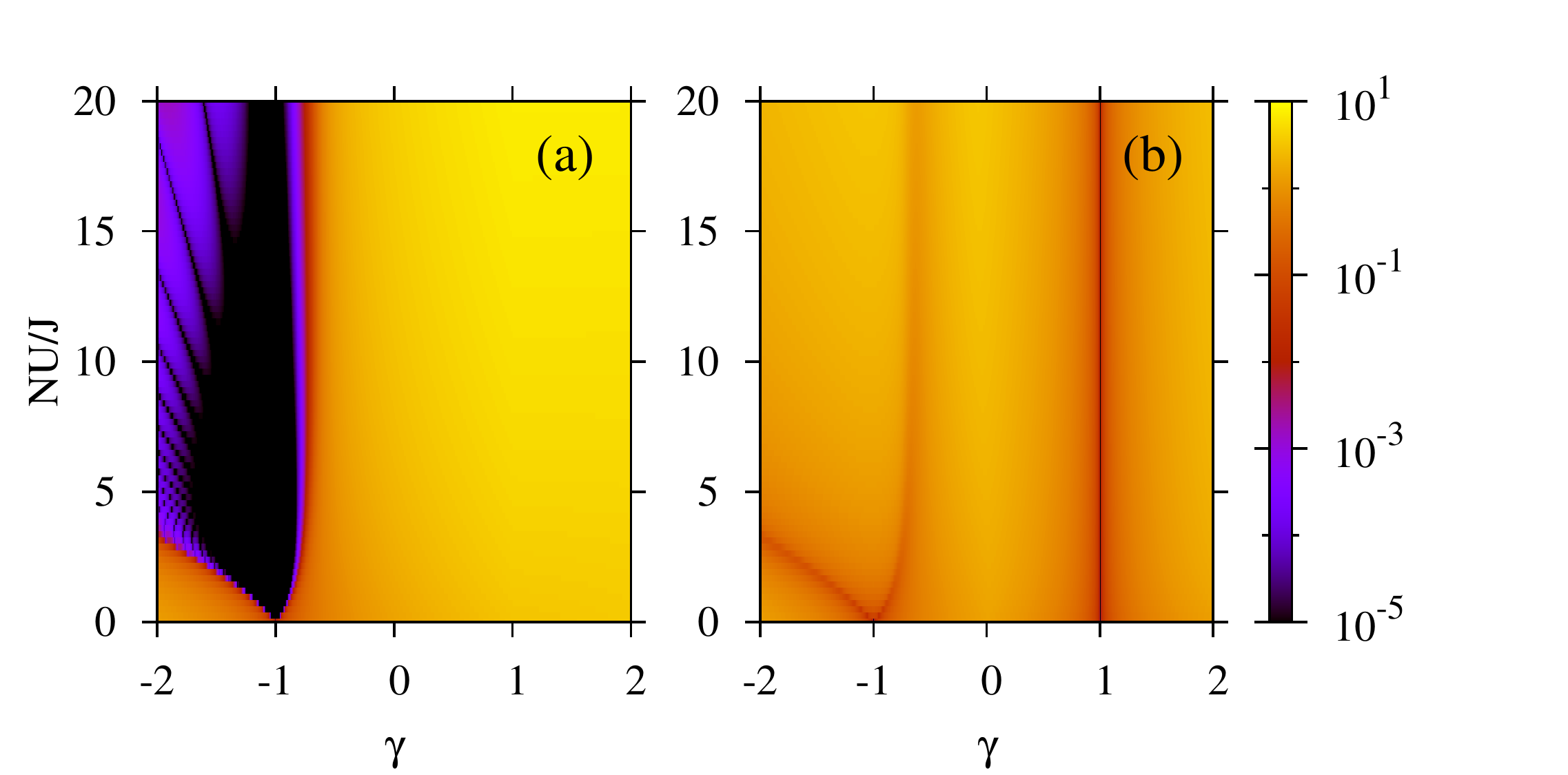}
\caption{(a) Energy gap between the ground state and the first excited state. 
(b) Energy difference between the second and first excited states. $N=48$.}
\label{ene24}
\end{figure}

The model presented in Sec.~\ref{sec:sm} predicts a number of 
degeneracies in the many-body energy spectrum of the system. 
Indeed, Eq.~(\ref{eq:deg}) predicts $N/3$ energy crossings for both 
$\gamma<-1$ and $\gamma>1$. As explained above, these predictions 
are expected to hold for $\gamma\lesssim -1$ and are indeed found 
in the exact many-body spectrum as seen in Fig.~\ref{ene24}. 
Varying the value of $k$ in Eq.~(\ref{eq:deg}) from 0 to $N/3$ one 
obtains the large $|\gamma|$ behavior of the different 
lines of zero gap in Fig.~\ref{ene24} (a). Similarly, albeit not 
shown in the figure, for $\gamma\gg1$ the corresponding 
gapless lines are also found in the system. 

For $\gamma>0$, also interesting is the closing of the energy gap between 
the first and second excited states for $\gamma=1$ (see the vertical line 
in the right panel of Figs.~\ref{ene24}). It corresponds to the degeneracy 
between vortex and antivortex states studied in Ref.~\cite{Lee2006}, whose 
wavefunctions have been previously obtained in the non-interacting case, 
see  Eq.~(\ref{vortexantivortex}). As the interaction is increased the 
degeneracy between the corresponding two states remains.

\subsection{Entanglement spectrum and entanglement entropy}

The many-body ground state has been found to be fragmented in the 
vicinity of $\gamma=-1$ and mostly condensed otherwise, for 
$NU/J \simeq 1$. Besides fragmentation, the three-site configuration 
considered allows one to study the onset of entanglement and correlations 
among the three different sites depending on the specific values 
of the parameters. To characterise the spatial entanglement properties 
we will use the Schmidt gap and the entanglement von Neumann entropy 
defined in Sect.~\ref{sec:ent}. 

\begin{figure}[t]
\centering
\includegraphics[width=0.9\columnwidth]{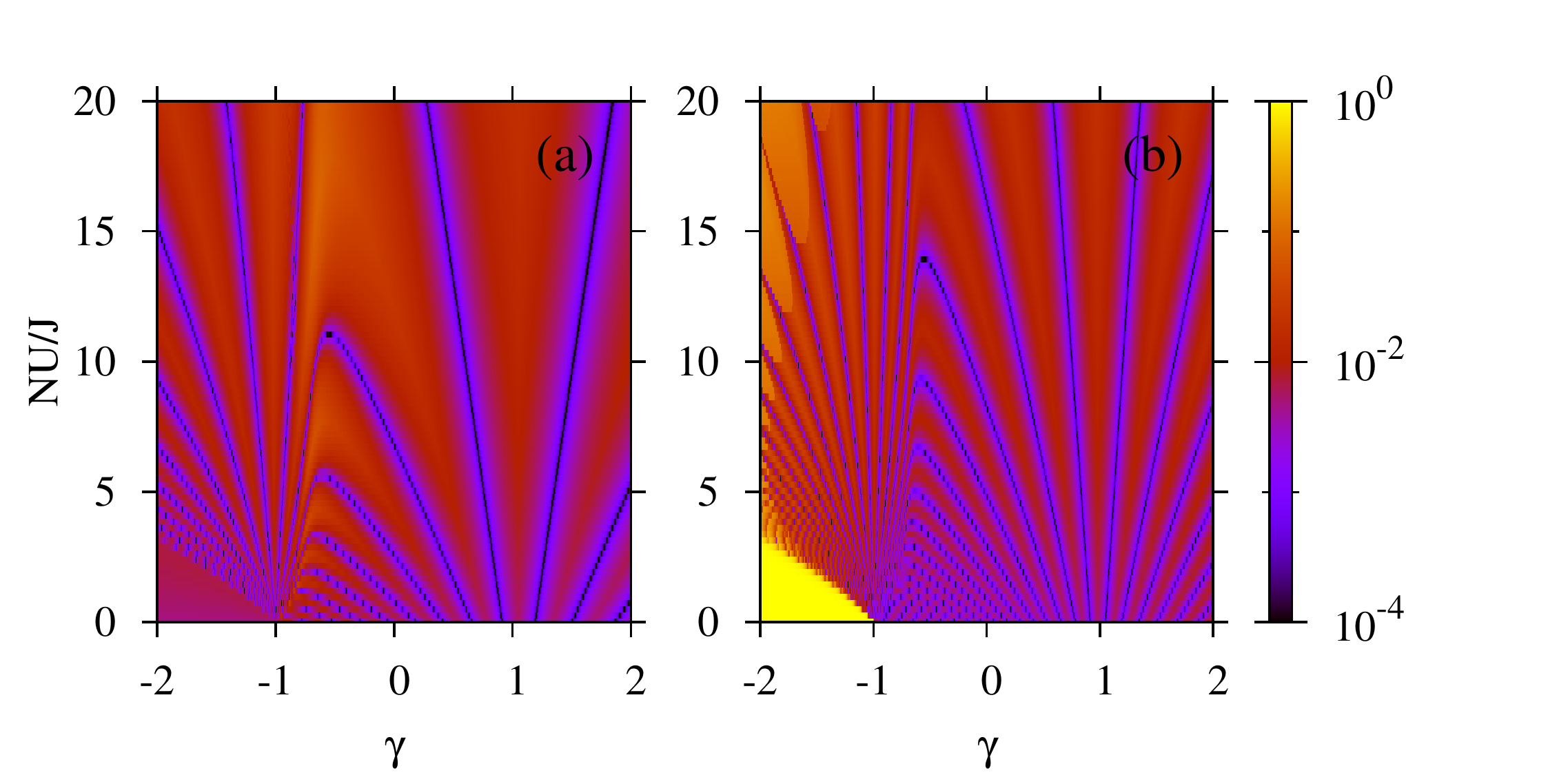}
\caption{Schmidt gap as a function of ($\gamma,\,N U/J$) for $N=48$. 
(a) Between site 1 and the subsystem formed by sites 2 and 3. 
(b) Between site 2 and the subsystem formed by sites 1 and 3. }
\label{schmidt}
\end{figure}

Due to the structure of our system, with three sites where two of them, 
1 and 3, are essentially equivalent, one can consider two bipartite 
splittings. The first one corresponds to subsystem 1 where sites 2 
and 3 have been traced off $(1,23)$. And second, subsystem 2 where subsystems 
1 and 3 have been traced off $(2,13)$.

\begin{figure}[t]
\centering
\includegraphics[width=.7\columnwidth]{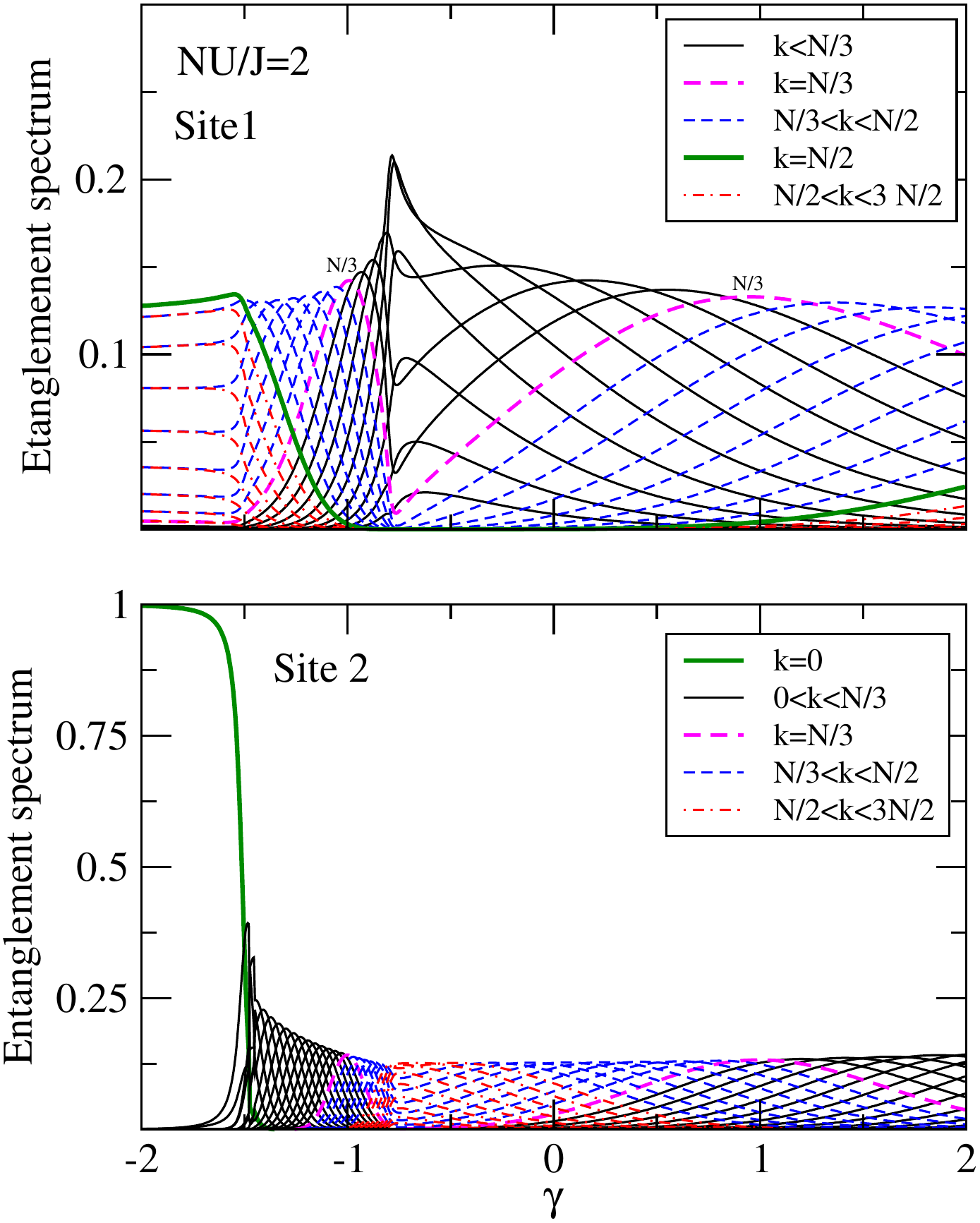}
\caption{Entanglement spectrum of the reduced density matrix of mode 
$1$ and mode 2, upper and lower panel, respectively. The spectrum is 
coloured according to the corresponding most probable occupation of the 
mode singlelling out some relevant cases, e.g. empty mode $k=0$, half 
occupation $k=N/2$, etc. Both are computed for $NU/J=2$ and $N=48$. }
\label{crossing}
\end{figure}

The Schmidt gap corresponding to both bipartite splittings in the 
$(\gamma, NU/J)$ diagram is presented in Fig.~\ref{schmidt}. The first notable 
feature is that in both bipartite splittings we observe two fan-like 
radial structures, with straight lines converging to $(-1,0)$ and $(1,c)$ (with 
$c$ a constant to be determined later). This structure, similar to the 
one discussed in Ref.~\cite{Gallemi2013}, represents the crossings of 
the first two values of the entanglement spectrum. To better 
understand the structure, in Fig.~\ref{crossing} we plot the entanglement 
spectrum ($N+1$ coefficients) for both bipartite splittings for a fixed 
value of $NU/J=2$. For $\gamma<-1$, each zero of the Schmidt gap shown 
in Fig.~\ref{schmidt} implies a variation of one unit of the most probable population 
of the untraced mode. This can be understood from the expression 
in Eq.~(\ref{rho_i}), where the reduced density matrix is shown to be diagonal 
in the Fock basis of the untraced mode. 

The simple large $|\gamma|$ model explained in section~\ref{sec:sm} also 
explains qualitatively the observed behaviour. The degeneracies in the 
energy spectrum obtained there correspond to many-body ground states 
in which the most likely value of the population in site 2 goes from 
0 to $N$/3 and the corresponding one of sites 1 and 3 from $N/2$ to $N/3$. 
We have considered $N=48$, see Fig.~\ref{schmidt}, and thus we find 16 crossings in the 
left part of the figure, with the most likely population of site 1 
going from $N/2=24$ in the leftmost case to $N/3=16$ at $\gamma=-1$. 
As explained in the previous section, the perturbative model captures 
the physics also for $\gamma>1$. As we can see in the figure, at 
$\gamma=1$ the most likely population of site 1 is again $N/3=16$ and 
it keeps increasing as $\gamma$ is increased. 

\begin{figure}[t]
\centering
\includegraphics[width=.9\columnwidth]{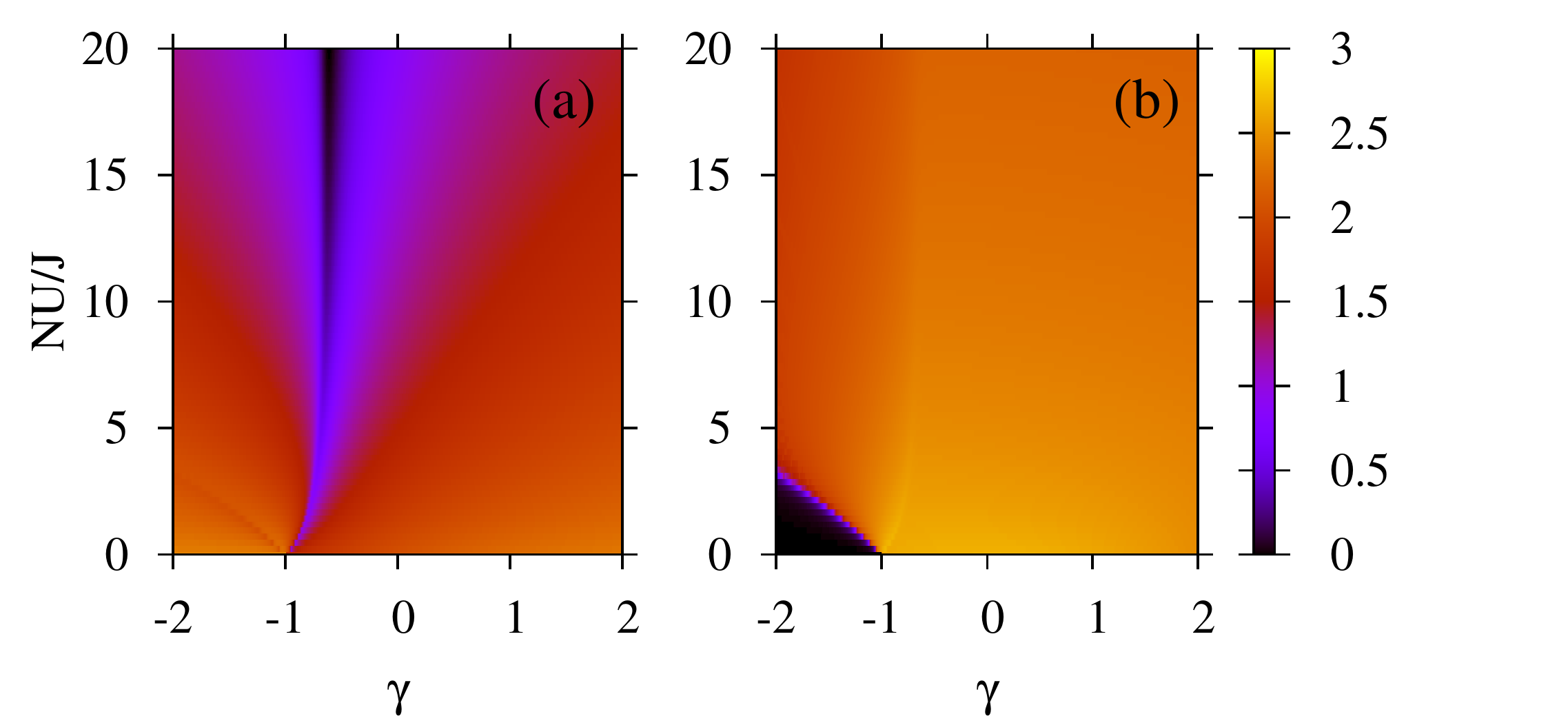}
\caption{Entropy as a function of ($\gamma,\,N U/J$) for $N=48$. 
(a) Bipartite von Neumann entropy between site 1 and the 
subsystem formed by sites 2 and 3. (b) Bipartite von Neumann 
entropy between site 2 and the subsystem formed by sites 1 and 3.}
\label{entropy}
\end{figure}

These crossings described in the entanglement spectrum, clearly visible in 
Fig.~\ref{crossing}, are one of the main signatures of a crossover between 
two phases, whereas in a quantum phase transition all the Schmidt coefficients 
degenerate to the same value in the thermodynamic limit 
($N\rightarrow\infty$)~\cite{DeChiara2012}. In our case, it is clear that at 
$(\gamma=-1,NU/J=0)$ all the Schmidt coefficients degenerate even for finite $N$. 
This is due to the fact that we already have a single-particle degeneracy.

Certain limiting cases are easily interpreted. For instance, for 
$\gamma\ll -1$, mode 2 is essentially unpopulated and decoupled 
from modes 1 and 3. This reflects in a singly populated entanglement 
spectrum for $NU/J=2$ in Fig.~\ref{crossing} (lower panel). In 
this regime, the system was shown to be condensed in Fig.~\ref{frag}, 
on the single-particle state $(|1,0,0\rangle+|0,0,1\rangle)/\sqrt{2}$, 
which spatially entangles modes 1 and 3, as seen in Fig.~\ref{crossing} 
(upper panel). In the $\gamma \gg1$ limit the situation is exactly the same, but 
as explained above, this limit is achieved in practice for much larger
values of $|\gamma|$ than in the $\gamma<0$ case. In the vicinity of 
$\gamma=0$ the three single-particle states quasidegenerate. This makes 
that for relatively low interactions, as in Fig.~\ref{crossing}, the many-body 
ground state starts populating the Fock states around $N/3$ approaching 
the Mott insulator phase. 

These features, described on the full entanglement spectrum, reflect 
directly on the corresponding von Neumann entropies, see 
Fig.~\ref{entropy}. For instance, the fact that the system approaches the 
Mott regime for relatively low values of the interaction in the vicinity 
of $\gamma=0$ reflects almost zero value of the von Neumann entropy for mode 
1, which starts to decouple from the other modes. For $\gamma<-1$ and 
small values of the interaction, the system empties mode 2 and decouples
it from the other two modes, see Fig.~\ref{entropy} (right panel).

\subsection{Quantum phase transition for attractive interactions}

\begin{figure}[t]
\centering
\includegraphics[width=0.9\columnwidth]{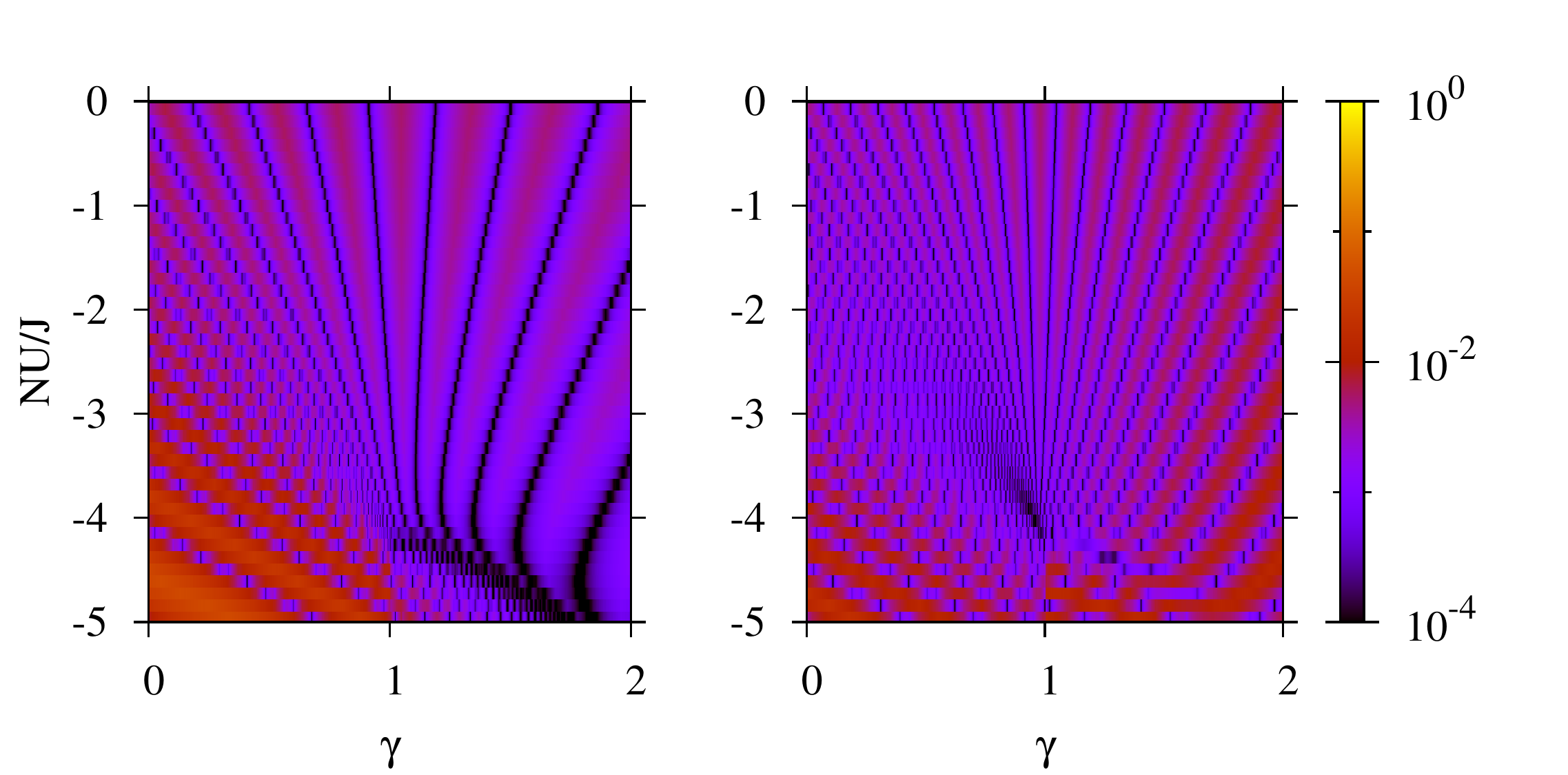}
\caption{Schmidt gap as a function of ($\gamma,\,N U/J$) for $N=48$. 
(a) Between site 1 and the subsystem formed by sites 2 and 3. 
(b) Between site 2 and the subsystem formed by sites 1 and 3. This figure 
is similar to Fig.~\ref{schmidt} but we extend to attractive interactions 
to show the phase transition described in the text.  }
\label{schmidtfocus}
\end{figure}

Similarly to the two-well system, a quantum phase transition can be 
expected for $\gamma=1$ for attractive 
interactions~\cite{Cirac1998,JuliaDiaz2010a}. In this case, the three sites 
are completely equivalent. Increasing the attractive atom-atom 
interactions, the system will minimise energy by clustering the atoms 
in a single site. In absence of any spatial bias, the ground 
state of the system will approach the Schr\"odinger cat-like state, 
\begin{equation}
|\Psi_{cs}\rangle={1\over \sqrt{3}}( |N,0,0\rangle+|0,N,0\rangle+|0,0,N\rangle) \,.
\label{eq:cs}
\end{equation} 
Analogously to the two-site case, this many-body state is not gapped and is 
quasidegenerate with its first two excitations. In the thermodynamic limit, 
the transition between the non-interacting state, Eq.~(\ref{coh3}), and 
the cat-like state Eq.~(\ref{eq:cs}) goes through a transition point at 
a finite value of $NU/J$. This transition reflects in the behaviour 
of the Schmidt gap of the system. As seen in Fig.~\ref{schmidt}, several  
zero-Schmidt-gap straight lines tend to converge on a point in the attractive 
interaction regime on the $\gamma=1$ line. In Fig.~\ref{schmidtfocus} we 
extend the range of parameters to the attractive region and certainly the 
lines seem to converge at $\gamma=1$ and $-9/2<NU/J<-4$, which is where the 
authors of Ref.~\cite{Dell'Anna2013} predicted the existence of a quantum phase 
transition. Notably, in contrast with the former phase transition described at $\gamma=-1$, 
this phase transition is only present in the thermodynamic limit. 

\section{Detailed analysis of the $\gamma=-1$ case}
\label{gam1}

As already shown in Figures ~\ref{frag} and~\ref{cfrag}, the $\gamma = -1$ ground state is found to be 
bifragmented, with $p_1=p_2= 1/2$, in the non-interacting limit ($NU/J =0$). To a 
good approximation it remains bifragmented for finite but small ($NU/J\simeq 1$) 
interactions. As $NU/J $ increases further the system approaches the Fock regime 
and the ground state tends to the well known trifragmented configuration with 
$p_1=p_2=p_3 = 1/3$. We will here instead focus on the description of the special 
bifragmented states obtained when  $NU/J$ is non-vanishing but small.

At $\gamma=-1$ the single 
particle spectrum, see Fig.~\ref{singlep}, has two degenerate eigenvalues, 
$E_S=E_A=-J$ and $E_e=2J$. A convenient choice for their eigenvectors is, 
\beqa
|S \rangle &=&  \frac{1}{\sqrt{6}} ( |1,0,0\rangle+ 2 |0,1,0\rangle + |0,0,1\rangle)\nonumber\\
|A \rangle &=&  \frac{1}{\sqrt{2}} ( |1,0,0\rangle -  |0,0,1\rangle)\nonumber\\
|e \rangle &=& \frac{1}{\sqrt{3}} ( |1,0,0\rangle - |0,1,0\rangle + |0,0,1\rangle)\,,
\label{eq:t5}
\eeqa
so that, under the exchange of sites $1 \leftrightarrow 3$, $|S \rangle $ is symmetric and $|A \rangle$ 
is antisymmetric. $|e \rangle$ corresponds to the ``excited'' single-particle 
configuration, $E_e-E_S = 3 J$. To these states we associate creation operators 
\beqa
{\hat S}^{\dag} &\equiv& \frac{1}{\sqrt{6}} \left(  \hat{a}_1^{\dag} + 2 \hat{a_2}^{\dag} + \hat{a_3}^{\dag}\right) \nonumber \\
{\hat A}^{\dag} &\equiv& \frac{1}{\sqrt{2}} \left( \hat{a}_1^{\dag} - \hat{a}_3^{\dag}\right) \nonumber \\
{\hat e}^{\dag} &\equiv& \frac{1}{\sqrt{3}}  \left( \hat{a}_1^{\dag} - \hat{a}_1^{\dag} +  \hat{a}_3^{\dag}\right)\,,
\label{eq:t15}
\eeqa
and their hermitian conjugates, ${\hat S}, {\hat A}$ and ${\hat e}$, act as annihilation 
operators. It is immediate that
\beqa
[ {\hat A} , {\hat A}^{\dag}] &=& 1 \quad , \quad [{\hat S} , {\hat S}^{\dag}] = 1 \quad , \quad [ {\hat e} , {\hat e}^{\dag}] = 1  \ .
\label{eq:t16}
\eeqa
Formally, it is a matter of convenience to use modes 1, 2, and 3 or $S$, $A$ 
and $e$ to construct the single-particle basis.\\ 
To understand the onset of bifragmentation we will now develop an 
approximate model in the following way: we will only use the subspace generated 
by the ${\hat A}^{\dag}$ and ${\hat S}^{\dag}$ modes to describe the lowest lying 
many body states, in particular the ground states. To do so, inverting Eq.~(\ref{eq:t15}) 
one finds $\hat{a}^{\dag}_i$ in terms of ${\hat A}^{\dag}, {\hat S}^{\dag}$ 
and ${\hat e}^{\dag}$ but since the latter do not operate in the chosen subspace, in our approximation, 
\beqa
\hat{a}_1^{\dag} &=&  \frac{1}{\sqrt{2}} {\hat A}^{\dag} + \frac{1}{\sqrt{6}} {\hat S}^{\dag} \nonumber \\
\hat{a}_2^{\dag} &=& \sqrt{\frac{2}{3}} {\hat S}^{\dag} \nonumber \\
\hat{a}_3^{\dag} &=& - \frac{1}{\sqrt{2}} {\hat A}^{\dag} + \frac{1}{\sqrt{6}} {\hat S}^{\dag} \ . \nonumber \\ 
\label{eq:t24}
 \eeqa
Within that subspace the interaction Hamiltonian reads, 
\beqa
\tilde{H}_U &=&  \sum_{i=1}^3 (\hat{a}_i^{\dag})^2 \hat{a}_i^2 = \frac{1}{2} ({\hat A}^{\dag})^2 {\hat A}^2 
+ \frac{1}{2} ({\hat S}^{\dag})^2 {\hat S}^2 \nonumber 
\\ &+& \frac{2}{3} {\hat S}^{\dag} {\hat A}^{\dag} {\hat S}  {\hat A} 
+ \frac{1}{6} \left(  ({\hat S}^{\dag})^2 {\hat A}^2 + ({\hat A}^{\dag})^2 {\hat S}^2 \right)  \ ,
\label{eq:t25}
\eeqa
where for simplicity we have omitted the $U/2$ factor and the terms linear in $N$. Introducing two new modes, 
\beqa
\hat{a}^\dagger_{hv1}  &=& {1\over\sqrt{2}} ( \hat{S}^\dagger+ i \hat{A}^\dagger)\nonumber \\
\hat{a}^\dagger_{hv2}  &=& {1\over\sqrt{2}} ( \hat{S}^\dagger- i \hat{A}^\dagger)
\label{eq:voo1}
\eeqa
the $\tilde{H}_U$ Hamiltonian becomes, 
\beqa
\tilde{H}_U = {1\over 3} \left( \hat{N}_{hv1}^2 +  \hat{N}_{hv2}^2 + 4 \hat{N}_{hv1}  \hat{N}_{hv2} - {\hat N}_{hv1} - {\hat N}_{hv2} \right)
\label{eq:hhu}
\eeqa
\begin{figure}[t]
\centering
\includegraphics[width=0.8\columnwidth]{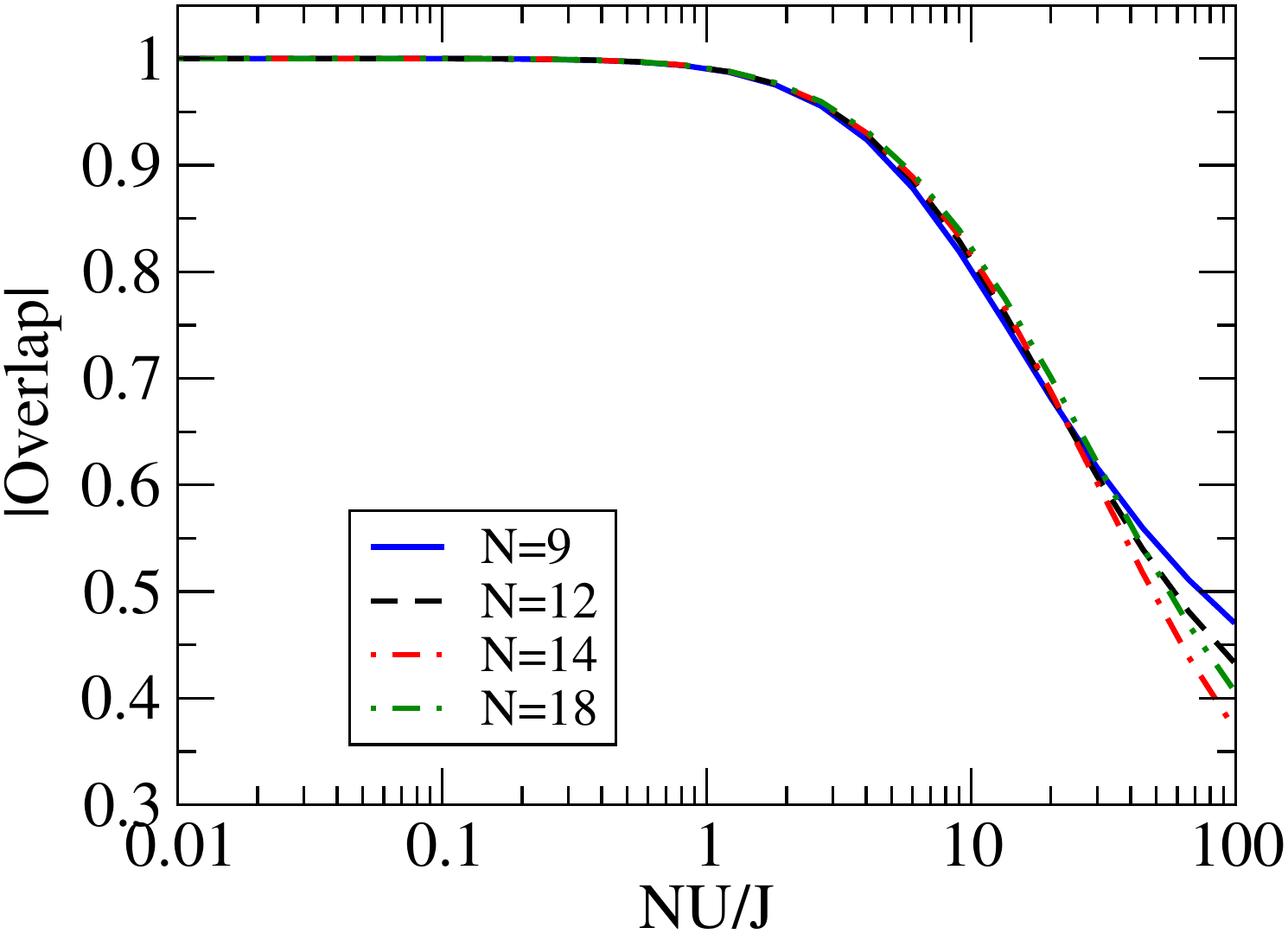}
\caption{Overlap between the exact two-fold quasidegenerate 
groundstate manifold, $|\Phi_j\rangle$ and the two $k=0$ states in 
Eq.~(\ref{ivcat}). The overlap is computed as $|{\rm det}{\cal O}|$, 
with ${\cal O}_{i,j} = \langle \Psi_0^i | \Phi_j\rangle$, $i=+,-$ and $j=1,2$. 
Note this overlap is 1 if the two sets of vectors span the same 
manifold. }
\label{figseciv}
\end{figure}
where $\hat{N}_{hv1}= \hat{a}^\dagger_{hv1} \hat{a}_{hv1}$ and 
$\hat{N}_{hv2}= \hat{a}^\dagger_{hv2} \hat{a}_{hv2}$. From Eq.~(\ref{eq:hhu}) 
it is clear that the Fock states built from the new modes
\beqa
|N_{hv1}, N_{hv2}\rangle &\equiv& \frac{1}{\sqrt{N_{hv1}! N_{hv2}!}} \ 
({\hat a}^{\dag}_{hv1} )^{N_{hv1}} \ ({\hat a}^{\dag}_{hv2})^{N_{hv2}} |{\rm vac}\rangle 
\eeqa
are the eigenvectors of ${\tilde H}_U$, with an obvious expression for the eigenvalues. 
Since $N_{hv1} + N_{hv2} =N$, it will be more clarifying to change notations to $N_{hv1} =k$ 
and $N_{hv2}= N-k$. Then  the spectrum becomes
\beq
{\tilde E}_U (k) = \frac{1}{3} \left( k^2 +(N-k)^2 + 4 k (N-k) -N \right)  \ ,
\eeq
which is degenerate in pairs, $(k, N-k)$, except the topmost energy when $N$ is odd. 
In particular, the ground state is two-fold degenerate.

\begin{figure}[t]
\centering
\includegraphics[width=0.8\columnwidth,angle=-0]{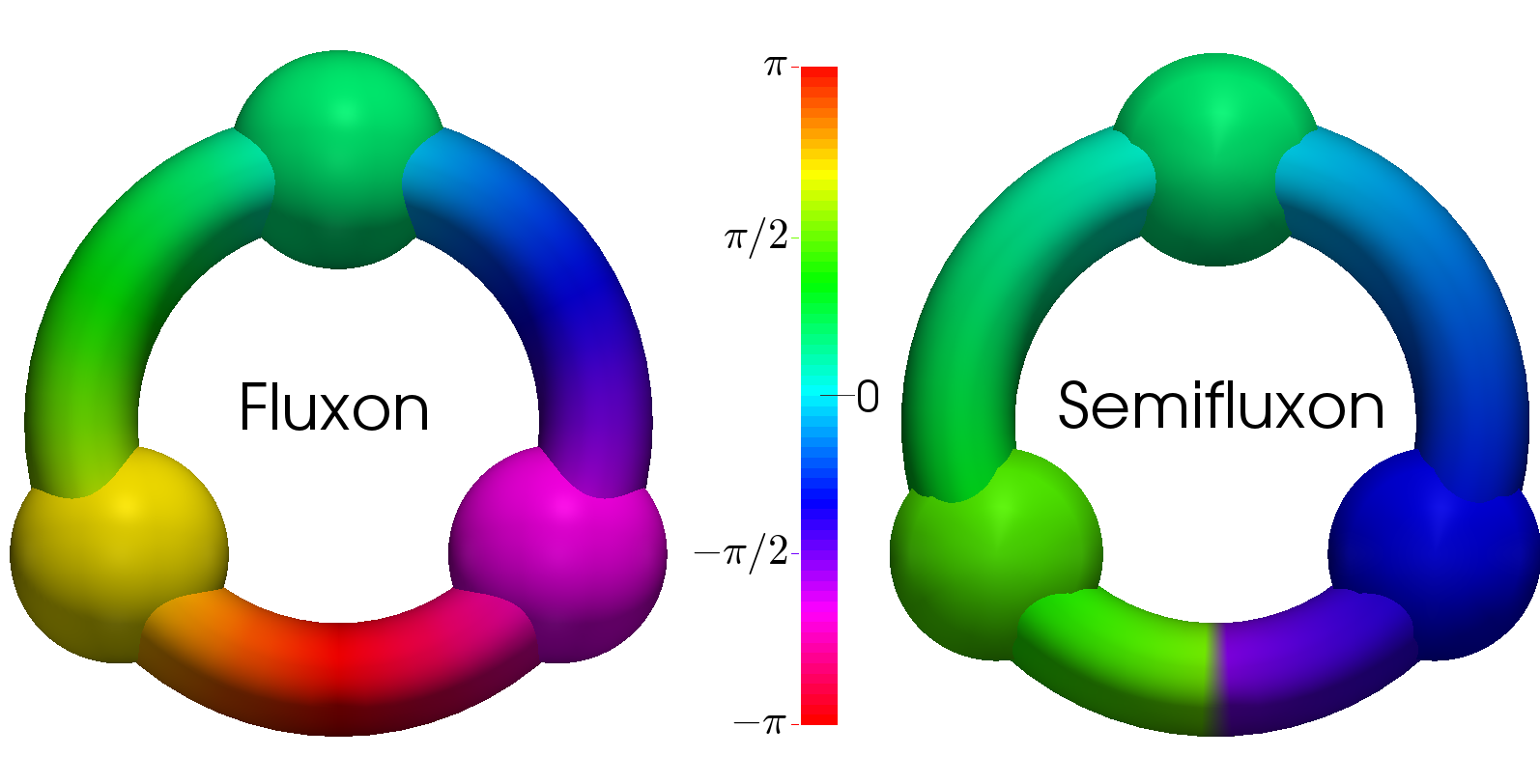}
\caption{Quantum phase in each site, color value, for the fluxon 
Eq.~(\ref{vortexantivortex}) and the semifluxon~ Eq.~(\ref{fvs}) 
states. The fluxon is the usual vortex state, with a full $2\pi$ variation 
of the phase around the triangle. In the semifluxon, the phase varies 
by $\pi$ as we go along the triangle. 
\label{fv}}
\end{figure}

This approximation turns out to be very accurate for small interactions. The 
effect of including the $|e\rangle$ manifold, breaks the degeneracy but does 
not promote any of the 
states. The lowest many body states are then well approximated by, 
\beqa
|\Psi_{k}^{(\pm)}\rangle &=& {1\over\sqrt{2}} 
\left[ |k, N-k\rangle \pm |N-k, k \rangle \right]  
\nonumber\\
&&{\rm with}\ k=0, \dots, N
\label{ivcat}
\eeqa
where the $\pm$ sign labels the two states which would be degenerate in energy 
in absence of coupling with the $|e\rangle$'s. The $\Psi_k^{(\pm)}$ are obviously 
bifragmented, and have $p_1=p_2=1/2$.  Furthermore, the ground state is a cat-state 
with all atoms in one or the other of the modes in Eq.~(\ref{eq:voo1}). These 
$a^{\dag}_{hv}$ can be considered creation operators of discrete semifluxon states ~\cite{fv}
because, combining Eqs.~(\ref{eq:t15}) and Eq.~(\ref{eq:voo1}), they can be written as 
\beqa
\hat{a}^\dagger_{hv1} &=& {1\over\sqrt{3}} \left( 
e^{i\pi/3} \hat{a}_1^\dagger 
+ \hat{a}_2^\dagger \ 
+ e^{-i\pi/3} \hat{a}_3^\dagger \right)\nonumber\\
\hat{a}^\dagger_{hv2} &=& {1\over\sqrt{3}} \left( 
e^{-i\pi/3} \hat{a}_1^\dagger 
+ \hat{a}_2^\dagger 
+ e^{i\pi/3} \hat{a}_3^\dagger \ \right)\ .
\label{fvs}
\eeqa
A semifluxon is a quantum state which carries half the quantum flux of 
the vortex states of Eq.~(\ref{vortexantivortex}). In our case, we have a 
discrete version, as going around the triangle the quantum phase grows from 
0 to $\pi$, with the phase jump of $\pi$ imposed by $\gamma=-1$. In 
Fig. \ref{fv} we depict the phase structure of the discrete semifluxon 
state, compared to the usual vortex, Eq.~(\ref{vortexantivortex}) 
(referred to as ``fluxon'').

In the full three-mode space, as said above, the degeneracy of each 
doublet $\Psi^{(\pm)}_k$, splits and therefore the ground state 
should be a NOON state of both semifluxon states. In 
Fig.~\ref{figseciv} we depict the overlap between the exact 
ground state and first excited state and the corresponding $k=0$, 
$\pm$ states in Eq.~(\ref{ivcat}). The model is seen to be fairly 
accurate up to $NU/J\simeq 10$. In an experimental measurement, 
one should thus find a fractional flux in one or the other direction.

\section{Summary and conclusions}
\label{sec:con}

A simple configuration consisting of three sites with a single 
tunable link has been shown to exhibit a large variety of quantum 
many-body properties. An important novelty of the proposed scheme 
is that we have considered cases in which the tunnelling rate 
in one of the links is either zero-phase 
or $\pi$-phase tunnelling. In both cases, we have performed diagonalisations 
for finite number of particles, up to $N=48$, and have examined the 
many-body properties of the systems as a function of the atom-atom 
interaction and the tunable tunnelling rates. 

By varying the tunnelling in one link, by means of the parameter 
$\gamma$, which is within reach experimentally as explained in the 
introduction, one explores configurations ranging from the colinear 
one $\gamma=0$, the fully symmetric one $\gamma=1$, to the symmetric 
$\pi$-phase one, $\gamma=-1$. In the first two cases, the many-body 
ground state for small interactions $NU/J \sim 1$ is highly condensed, 
with one of the three eigenvalues of the one-body density matrix 
clearly scaling with the total number of particles. As we approach 
the $\gamma=-1$ point the system departs from condensation and becomes 
bifragmented for small interactions. This is partly a 
consequence of the degeneracy in the single-particle spectrum for $\gamma=-1$. 

Two quantum phase transitions are present in this system. The first one 
takes place in the symmetric configuration $\gamma=1$, but only for 
attractive interactions, which therefore makes it difficult to explore
experimentally. This phase transition is reminiscent of the one present 
in other few-site models, e.g. bosonic Josephson junctions, and marks 
the transition to Schr\"odinger cat-like ground states in the 
spectrum~\cite{Cirac1998,JuliaDiaz2010a,rav14}. Interestingly, 
for $\gamma=-1$ we find a second quantum phase transition, which 
takes place in absence of interactions and which has clear consequences 
for small repulsive atom-atom interactions. This phase transition has 
been characterised by the behaviour of the entanglement spectrum for the 
two possible independent bipartite splittings as we approach the transition 
point~\cite{DeChiara2012}. At the transition the entanglement spectrum 
degenerates and the corresponding von Neumann entropies exhibit a 
maximum. Interactions do not wash out the main features of this transition, 
which becomes essentially a crossover, which has clear consequences 
at finite interaction, such as the bifragmentation discussed above. 

Finally, for the $\gamma=-1$ case and small interactions, we have 
been able to derive an approximate many-body Hamiltonian, 
which describes the low-energy excitations as excitations of semifluxons. 
The bifragmentation is readily explained in this limit, 
and the ground state of the system is found to be a macroscopic 
Schr\"odinger cat state of discrete semifluxons with opposite 
currents.

\section*{Acknowledgements}
We acknowledge financial support from the Spanish MINECO 
(FIS2011-28617-C02-01 and FIS2011-24154) and the European 
Regional development Fund, Generalitat de Catalunya Grant 
No. 2014 SGR 401. A.G. is supported by Generalitat de 
Catalunya Grant FI-DGR 2014. B.J-D. is supported by the 
Ram\'on y Cajal MINECO program.

 \appendix

\section{Diagonal structure of reduced density matrices}
\label{app:diagonality}

The many-body ground state of a Bose-Hubbard Hamiltonian of $N$ atoms in
a triple-well potential,  can be 
expressed as 
$$|\Psi\rangle=\sum_{k,l} C_{k,l} \, |k,l,N-k-l\rangle \,,$$ 
where the Fock basis of the system can be written as a product state
of the reduced Fock basis  $ \{ |k \rangle_j \}$ for each subsystem $j=1,2,3$:
$$ |k,l,N-k-l\rangle = |k\rangle_1 \otimes |l\rangle_2 \otimes |N-k-l\rangle_3 \,.$$
The reduced density matrix of subsystem 1 can be computed as

\begin{eqnarray}
 \hat{\rho}_1 &=& \sum_{m,n} {_2} \langle m |{_3} \langle n | \left( \sum_{k,l} \, C_{k,l} |k,l,N-k-l\rangle  
 \sum_{k',l'} C^{*}_{k',l'}  \langle k',l',N-k'-l'| \right) | m \rangle _2 | n \rangle _3 \nonumber \\
 &=&\sum_{m,n} \sum_{k,k'} \sum_{l,l'}  \delta_{k,m} \, \delta_{N-k-l,n} \, \delta_{N-k'-l',n} \, \delta_{k',m}  \,
 C^{*}_{k',l'} C_{k,l} \,\, {_1}|l\rangle \langle l'|{_1}
\nonumber\\
&=&\sum_{m,n} |C_{m,n}|^{2} \, |N-m-n\rangle \langle N-m-n| \,.
\end{eqnarray}
From this expression one can see that 
the reduced density matrix of one subsystem $i$ is diagonal in the reduced Fock basis of the corresponding mode
$ \{ |k \rangle \}\equiv \{ |k \rangle_i \} \ $:
\begin{equation}
\hat{\rho}_i = \sum_{k=0}^N \lambda_{k}^i \, |k\rangle \langle k|\,,
\label{rho_i}
\end{equation}
with $\{ \lambda_k^i \}$ the Schmidt coefficients that contain the information 
concerning the entanglement properties of subsystem $i$ with the rest of the system.


\end{document}